\documentclass[10pt]{article}

\usepackage{graphicx}              
\usepackage{amsmath}               
\usepackage{amssymb, amsbsy, latexsym} 
\usepackage{amsfonts}              
\usepackage[mathscr]{eucal}     
\usepackage{amsthm}                
\usepackage{multirow}     
\usepackage{txfonts}

\DeclareMathSymbol{\C}{\mathbin}{AMSb}{"43}
\DeclareSymbolFont{AMSb}{U}{msb}{m}{n}

\makeatletter
\@addtoreset{equation}{section}
\makeatother

\newtheorem{Theorem}{Theorem}[section]
\newtheorem{Definition}{Definition}[section]

\def\be{\begin{eqnarray}}
\def\ee{\end{eqnarray}}

\newcommand{\ch}{\mathcal H}
\newcommand{\ci}{\mathcal I}

  \newcommand{\Fd}{\mathfrak{D}}

  \newcommand{\Fn}{\mathfrak{N}}

\newcommand{\eps}{\epsilon}

\renewcommand{\l}{\lambda}

\renewcommand{\O}{\Omega}

\newcommand{\rmd}{\mathrm d}

\newcommand{\lt}{\left}
\newcommand{\rt}{\right}

\newcommand{\lag}{\left\langle}
\newcommand{\rag}{\right\rangle}

\newcommand{\M}{\hat{\mathbf{M}}}

\newcommand{\scrD}{\mathcal{D}}

\begin{document}

\title{{\sf On the Relation between Rigging Inner Product and Master 
Constraint Direct Integral Decomposition}}

\author{
{\sf Muxin Han$^{1,3}$}\thanks{{\sf 
mhan@aei.mpg.de}},\ \   
{\sf Thomas Thiemann$^{1,2,3}$}\thanks{{\sf 
thiemann@aei.mpg.de, 
tthiemann@perimeterinstitute.ca,thiemann@theorie3.physik.uni-erlangen.de}}\\
\\
{\sf $^1$ MPI f. Gravitationsphysik, Albert-Einstein-Institut,} \\
           {\sf Am M\"uhlenberg 1, 14476 Potsdam, Germany}\\
\\
{\sf $^2$ Perimeter Institute for Theoretical Physics,} \\
{\sf 31 Caroline Street N, Waterloo, ON N2L 2Y5, Canada}\\
\\
{\sf $^3$ Institut f. Theoretische Physik III, Universit\"{a}t 
Erlangen-N\"{u}rnberg} \\
{\sf Staudtstra{\ss}e 7, 91058 Erlangen, Germany}
}
\date{}
\maketitle

\thispagestyle{empty}

\begin{abstract} 
{\sf
Canonical quantisation of constrained systems with first class 
constraints via Dirac's operator constraint method proceeds by 
the thory of Rigged Hilbert spaces, sometimes also called Refined 
Algebraic Quantisation (RAQ). This method can work when the constraints 
form a Lie algebra. When the constraints only close with nontrivial
structure functions, the Rigging map can no longer be defined.

To overcome this obstacle, the Master Constraint Method has been 
proposed which replaces the individual constraints by a weighted sum of
absolute squares of the constraints. Now the direct integral 
decomposition methods (DID), 
which are closely related to Rigged Hilbert spaces, become available  
and have been successfully tested in various situations.

It is relatively straightforward to relate the Rigging Inner Product to
the path integral that one obtains via reduced phase space methods. 
However, for the Master Constraint this is not at all obvious. In this 
paper we find sufficient conditions under which such a relation can be 
established. Key to our analysis 
is the possibility to pass to equivalent, Abelian constraints, at least
locally in phase space. Then the Master Constraint DID for those Abelian 
constraints can be directly related to the Rigging Map and therefore has 
a path integral formulation.
}
\end{abstract}

\newpage 

\tableofcontents

\newpage

\section{Introduction}

The quantization of a constrained system is of profound interest, 
because the fundamental interactions in the physical world are described 
by theories with gauge symmetries.
The case of General Relativity is especially 
interesting and challenging, because its Hamiltonian is a linear 
combination of the first-class constraints, which means that the 
dynamics of GR is determined by the constraints and their gauge 
transformations.

There are many different approaches to quantize a constrained system 
(see \cite{HT}), one of which is canonical quantization which uses
the operator formalism. A 
traditional way to perform canonical quantization for a constrained
system is Dirac quantization \cite{dirac}. In 
Dirac quantization we first perform the 
quantization procedure disregarding the constraints and define 
a certain kinematical Hilbert space $\ch_{Kin}$, which provides a 
representation of the elementary variables and their canonical 
commutation relations. Then we quantize the classical first-class 
constraints $C_I$ as densely defined and closable 
operators $\hat{C}_I$ 
on the 
kinematical Hilbert space $\ch_{Kin}$. Once such a construction is 
finished, we should define the Quantum Constraint Equation 
\be
\hat{C}_I\Psi=0 \label{CPsi}
\ee
and solve it in general. The space of solutions equipped with a  
physical inner product defines the physical Hilbert space. Such a 
prescription is no problem when we consider the simplest case that there 
is only one single constraint $\hat{C}$, and that $\hat{C}$ is a 
self-adjoint 
operator with only pure point spectrum. It is because in this case, we 
only need to solve the eigenvalue equation $\hat{C}\Psi=0$ corresponding 
to 
the zero eigenvalue, and the space of solutions is a subspace of the 
kinematical Hilbert space. Therefore the physical inner product is 
the same as the kinematical inner product without ambiguity.  
The physical Hilbert space $\ch_{Phys}$ is identified as Hilbert 
subspace of the kinematical Hilbert space $\ch_{Kin}$ corresponding to 
the constraint kernel. However, the 
above naive prescription of Dirac quantization often fails to specify 
the physical Hilbert space for more complicated constrained systems. 
The complications may come from the following sources:

\begin{itemize}
\item The constrained system may possess several constraints $C_I$ 
$I\in\ci$ where $\ci$ is a (finite or infinite) index set. If we can 
represent all the constraints as operators $\hat{C}_I$, it is in general 
hard to solve all the constraints together and find the common solution 
spaces.

\item The first-class constraints $C_I$ form a constraint algebra with 
the Poisson commutation relation
\be
\lt\{C_I,C_J\rt\}=f_{IJ}^{\ \ K}C_K
\ee
where in general $f_{IJ}^{\ \ K}$ may be a function depending on the 
phase space variables ($f_{IJ}^{\ \ K}$ is called a structure function). 
The quantization of the constraints in this case may suffer from 
quantum anomalies, which results in the physical Hilbert space to have 
less
degrees of freedom than the classical theory.

\item Even when we don't have the above problems, e.g. even when we 
consider just a single self-adjoint constraint operator $\hat{C}$, there 
is still the problem about how to specify the physical inner product for 
the solution space. The issue arises because the spectrum of the 
constraint operator 
$\hat{C}$ in general is not only pure point, but can also have a
continuous part. If zero is contained in the continuous spectrum, 
the solutions of the quantum constraint equation Eq.(\ref{CPsi}) are in 
general not contained in the kinematical Hilbert space $\ch_{Kin}$. Thus 
the inner product of $\ch_{Kin}$ is not available for the 
definition of the physical inner product, because the solution space of 
the 
quantum constraint equation is not a subspace of the kinematical Hilbert 
space anymore. 

\end{itemize}
 
In this paper we consider two approaches that have been proposed to 
refine Dirac's 
quantization procedure and in 
order to (partially) solve the above problems: 

The first one is the so called, Refined Algebraic Quantization (RAQ) 
\cite{RAQ} programme. The RAQ prescription relaxes the condition that 
the 
solution of the constraint equations belongs to the kinematical 
Hilbert space. Solutions to the constraints are now elements 
of the algebraic dual 
$\Fd^\star_{Kin}$, that is, distributions on a dense domain 
$\Fd_{Kin}\subset\ch_{Kin}$, which 
is supposed to be invariant under all the $\hat{C}_I$ and 
$\hat{C}_I^\dagger$ (the constraint operator may not necessarily be
self-adjoint). So what we are looking are states 
$\Psi\in\Fd^\star$ such that:
\be
\Psi\lt[\hat{C}^\dagger_If\rt]:=\hat{C}'_I\Psi\lt[f\rt]=0,\ \ \ \ \ \ 
\forall f\in\Fd
\ee
The space of solutions is denoted by $\Fd^\star_{Phys}$. The physical 
Hilbert space will be a subspace of $\Fd^\star_{Phys}$. Eventually, 
$\Fd^\star_{Phys}$ will be the algebraic dual of a dense domain 
$\Fd_{Phys}\in\ch_{Phys}$, which is invariant under the algebra of 
operators corresponding to Dirac observables. Hence we obtain a Gel'fand 
triple:
\be
\Fd_{Phys}\hookrightarrow\ch_{Phys}\hookrightarrow \Fd^\star_{Phys}
\ee
A systematic construction of the physical Hilbert space is available if 
we have an anti-linear rigging map:
\be
\eta:\Fd_{Kin}\to\Fd_{Phys}^\star;\ f\mapsto\eta(f)
\ee
such that $\eta(f')[f]$ is a positive semi-definite sesquilinear form on 
$\Fd_{Kin}$ and such that 
for all the Dirac observables $\hat{O}$ on the kinematical Hilbert 
space, we have $\hat{O}'\eta(f)=\eta(\hat{O}f)$. 
If the quantum constraint algebra is generated by 
self-adjoint constraints $\hat{C}_I$ and their commutator algebra is a 
Lie 
algebra i.e. the structure functions are constant, then we can try to  
heuristically define the 
rigging 
map via the group averaging procedure:
\be
\eta(f):=\int\rmd\mu(t)\ <e^{it^I\hat{C}_I}f,.>
\ee
where $\rmd\mu$ is an invariant measure on the gauge group generated by 
the constraints, e.g. if the gauge group is a locally compact Lie group, 
$\rmd\mu$ can be chosen as the Haar measure. If we have obtained a 
rigging map $\eta$, the physical inner product is defined by the rigging 
inner product
\be
\lag\eta(f)|\eta(f')\rag_{Phys}:=\eta(f')[f],\ \forall f,f'\in\Fd_{Kin}.
\ee
Then a null space $\Fn\subset\Fd_{Phys}^\star$ is defined by 
$\lt\{\eta(f)\in\Fd_{Phys}^\star\ \big|\ ||\eta(f)||_{Phys}=0\ \rt\}$. 
Therefore 
\be
\Fd_{Phys}:=\eta\lt(\Fd_{Kin}\rt)/\Fn
\ee
The physical Hilbert space $\ch_{Phys}$ is defined by the completion 
of $\Fd_{Phys}$ with respect to the physical inner product. The above 
prescription of RAQ provides  an effective way to obtain the physical 
Hilbert space by quantizing a general first-class constrained system, 
whose constraint algebra has a Lie algebra structure and the quantum 
gauge transformations form a group such that group averaging can be 
applied. However, this prescription is not applicable to a 
constraint algebra with structure functions. 

The new idea put forward in \cite{link} is to exploit the 
Abelianization 
theorem \cite{HT} in order to adapt RAQ to the case with non trivial 
structure functions. 
The Abelianization theorem states that in general, all the 
first-class constraints can be abelianized at least locally in 
the phase space, i.e. there exists a family of constraints $\tilde{C}_I$ 
(locally) equivalent to the original family of constraints, such that 
$\{\tilde{C}_I,\tilde{C}_J\}=0$. 
If the Abelianized constraints $\tilde{C}_I$ can be quantized as 
self-adjoint operators without anomalies, that is, 
$[\tilde{C}_I,\tilde{C}_J]=0$, we obtain a 
quantum constraint algebra with Lie algebra structure and the 
quantum gauge transformations generated by them form an Abelian group. 
Thus we can use the group averaging technique to construct the rigging 
map and the physical Hilbert space as sketched above.

Another proposal is the Master Constraint Programme (MCP) 
\cite{master} and Direct Integral Decomposition (DID) 
\cite{DID,DID2}. The MCP modifies the prescription of Dirac quantization 
by introducing a so called, Master Constraint, which is 
classically defined by
\be
\textbf{M}:=\sum_{I,J\in\ci}{K}^{IJ}C_I{C}_J
\ee
for some real valued positive matrix ${K}^{IJ}$ which could even be a 
non trivial function on phase space. Classically one has 
$\textbf{M}=0$ if and only if $C_I=0$ for all $I\in\ci$. Also the Dirac 
observables can be defined purely in terms of M \cite{master}. Thus
$M$ is a classically equivalent starting point in order to encode the 
full set of constraints $C_I$. It is therefore conceivable that the 
quantized master constraint $\M$ can be used as an alternative tool 
in order to determine the physical Hilbert space in the situation that
group averaging with respect to the individual constraints is available
and that it extends RAQ to the situation with non trivial structure 
functions. This expectation has been verified in many non trivial 
examples \cite{DID,DID2}. 

An immediate technical advantage of the master 
constraint over the individual constraints is that, as 
a positive operator, the master constraint $\textbf{M}$ can be 
defined as a self-adjoint operator on $\mathcal{H}_{Kin}$ by employing 
the preferred 
Friedrich's self-joint extension \cite{Simon}. Moreover, if the 
kinematical 
Hilbert space is separable, the physical Hilbert space can be obtained 
via spectral theory, specifically 
Direct Integral Decomposition (DID). We first recall the general 
definition of the DID representation of the Hilbert space.

\begin{Definition} \label{DID}
Let $(X,{\cal B},\mu)$ be a separable topological measure space such 
that $X$ is  
$\sigma-$finite with respect to $\mu$ 
and let 
$x\mapsto {\cal H}_x$ be an assignment
of separable Hilbert spaces such that the function $x\mapsto N(x)$, 
where 
$N(x)$
is the countable dimension of ${\cal H}_x$, is measurable.
It follows that the sets $X_N=\{x\in X;\;N(x)=N\}$,
where $N$ denotes any countable cardinality, are measurable. Since 
Hilbert 
spaces
whose dimensions have the same cardinality are unitarily equivalent we 
may
identify all the ${\cal H}_x,\;N(x)=N$ with a single ${\cal 
H}_N=\mathbb{C}^N$
with standard $l_2$ inner product. 
We now consider maps
\begin{eqnarray}
\psi:\; X\to \prod_{x\in X} {\cal H}_x;\;\;
x\mapsto (\psi(x))_{x\in X} \label{0.12}
\end{eqnarray}
subject to the following two constraints: \\
1. The maps $x\mapsto <\psi(x),\psi(x)>_{{\cal H}_N}$ are measurable for
all $x\in X_N$ and all $\psi \in {\cal H}_N$. \\
2. If
\begin{eqnarray}
<\psi_1,\psi_2>:=\sum_N \int_{X_N}\; d\mu(x)
<\psi_1(x),\psi_2(x)>_{{\cal H}_N}\label{0.13}
\end{eqnarray}
then $<\psi,\psi> <\infty$.\\
The completion of the space of maps (\ref{0.12}) in the inner product
(\ref{0.13}) is called the direct integral of the ${\cal H}_x$ with
respect to $\mu$ and one writes
\begin{eqnarray}
{\cal H}^\oplus_{\mu,N}=\int_X^\oplus\;\rmd\mu(x)\; {\cal H}_x,\;\;\;
<\xi_1,\xi_2>=
\int_X\;\rmd\mu(x)\; <\xi_1(x),\xi_2(x)>_{{\cal H}_x}\label{0.14}
\end{eqnarray}
\end{Definition}

Here in our case, the spectral theorem for the self-adjoint master 
constraint $\M$ provides a natural DID representation of the kinematical 
Hilbert space $\ch_{Kin}$, where the topological measure space is the 
spectrum of the master constraint operator $\M$ and $\rmd\mu$ is the 
spectral measure. Then the physical Hilbert space is defined by the 
fiber Hilbert space $\ch_{x=0}$\footnote{Such a definition of the 
physical Hilbert space is in general ambiguous, there are some more 
physical prescriptions necessary to remove these ambiguities \cite{DID}. 
We will 
come back to 
this point in Section \ref{1}.}  

Notice that heuristically DID is nothing else than group averaging
for a single self -- adjoint constraint operator $\M$. The other 
advantage of the Master Constraint Programme is that there are no 
problems with anomalies as far as $\M$ itself is concerned since 
trivially $[\M,\M]=0$. Of course, if the individual constraints that 
constitute $\M$ are anomalous then $\M$ is expected to have trivial 
kernel and in this case one proposal is to subtract the corresponding 
spectral gap from $\M$, see \cite{master} for details. 

The master 
constraint rigging map is then heuristically defined  for any 
kinematical state $f\in\Fd_{Kin}$ via
\be
\tilde{\eta}(f):=\int\rmd t\ <e^{it\M}f,.>\label{mastergroup}
\ee
which also gives the physical inner product as a rigging inner product, 
and further gives the physical Hilbert space $\ch_{Phys}$.

Now we have three different approaches towards the physical Hilbert 
space 
of a general first-class constraint system. They are: 
\begin{enumerate}

\item The Direct Integral Decomposition (DID) using the master 
constraint,

\item The Refined Algebraic Quantization (RAQ) and the group averaging 
using the master constraint,

\item The Refined Algebraic Quantization (RAQ) and the group averaging 
using a set of Abelianized constraints.

\end{enumerate}

The immediate question to ask is: Are these three approaches 
equivalent? If not, which one gives the correct physical Hilbert space? 
For the examples discussed in \cite{DID2} it turned out that the DID 
approach using the master 
constraint always gave satisfactory results and 
to some extent is less ambiguous than the RAQ prescription. Moreover,
in \cite{DID} it was shown that RAQ with group averaging is in general 
inequivalent with DID, especially when zero is an eigenvalue 
embedded in the continuous spectrum in which case RAQ with group 
averaging sometimes leads to unsatifactory results.

The purpose of the present paper is to analyze in more detail the 
relations between the three prescriptions for the physical Hilbert 
space. It turns out that although the group averaging in the form of 
Eq.(\ref{mastergroup}) is inconsistent with the DID definition of the 
physical Hilbert space, a certain modification of the group averaging 
prescription Eq.(\ref{mastergroup}) does lead to consistency with 
the DID definition. More precisely, under certain technical 
assumptions, the modified group averaging technique captures precisely 
the absolutely continuous sector of the DID physical Hilbert space. 
The technical assumptions for establishing the consistency are fulfilled 
by all the physical models tested in \cite{DID2}. 

On the other hand, a 
similar modification of the group averaging prescription can also be 
done for the group averaging of the Abelianized constraints. It 
turns out again that under certain technical assumptions, the 
modified 
group averaging using the set of Abelianized constraints leads to the 
same result as the modified group averaging using a single master 
constraint for those Abelianized constraints. To summarize, under some 
assumptions which we spell out in detail in the course of this paper, 
the above 
three approaches for the physical Hilbert space are consistent among 
each other.
 
Our motivation for studying this questions arose from an important open 
question in Loop Quantum Gravity (LQG) \cite{book,rev}. LQG is a 
specific incarnation of the programme of canonical quantisation applied 
to General Relativity. It is a canonical quantum theory in terms of 
operators and Hilbert spaces. On the other hand, path integral 
techniques have been applied to LQG based on the kinematical Hilbert 
space underlying the canonical theory and resulted in what is called 
spin foam models \cite{spinfoam2}. While the two theories should both be 
quantisations of GR, the relation between the two is not at all obvious 
because in spin foam models one only uses the kinematical structure of 
LQG, the information about the quantum dynamics of the 
canonical theory \cite{QSD} is not obviously 
implemented in spin foam models which are formulated as 
(simplicity) constrained BF 
theories \cite{bf,plebanski}. In order to compare 
the canonical and spin foam approach it is natural to try to give a 
systematic path
integral derivation of spin foam models starting 
from the canonical theory,
which so far is missing entirely.

Now it is rather well known how to relate the group averaging map for
the individual constraints to the established reduced phase space path 
integral \cite{HT}, at least at a heuristic level. However, the 
constraints of GR are not of the kind to which group averaging 
techniques apply, since (in)famously they only close with non trivial 
structure functions which causes all sorts of technical problems (see 
e.g. the extensive discussion in \cite{AQGIV}). It
is for that reason that the Master Constraint
Programme was invented. However, the Master Constraint group averaging 
map is not obviously related to the path integral formulation of the 
individual constraints. The missing link between the path integral 
formulation and the Master constraint programme can be found by 
considering the intermediate step of group averaging the {\it 
Abelianized} constraints and the Master constraint for those {\it 
Abelianized} constraints. In \cite{link} we have sketched how 
one can directly relate the group averaging maps $\eta,\tilde{\eta}$ 
for these
Abelianised constraints and therefore has access to a path integral 
formulation directly from the Master constraint. In this paper we wish 
to study this relation mathematically 
more carefully.

One can rightfully ask whether all of this has any practical use as far 
as Quantum Gravity is concerned because the Abelianisation of 
constraints in field theories usually can be performed only at the 
price of giving up spatial locality. For instance, in pure gravity 
one can form four algebraically independent scalars out of the 3D 
Riemann
curvature and higher derivatives or polynomials thereof. In order to 
Abelianise the Hamiltonian and spatial diffeomorphism constraints of GR 
one needs to find a canonical transformation mapping to those scalars as 
configuration coordinates on phase space. It is clear that this 
involves inverting Laplacians. One then solves the 
constraints for the conjugate momenta of those scalars which provides 
the Abelianised constraints. This procedure is practically useless.
The idea therefore is to use suitable matter in order to avoid non 
locality which can be done \cite{BKR, phantom,tina} and in principle,
at least at a heuristic level, leads to a spin foam model, albeit 
necessarily with matter.\\      
\\
The present paper is organized as the follows:\\
\\
In section \ref{1}, we define a modified group averaging using a single 
self-adjoint master constraint operator, and prove under which 
circumstances such a group averaging gives the absolutely continuous 
sector of the DID physical Hilbert space.

In section \ref{2}, we define the modified group averaging using a set 
of self-adjoint Abelianized constraints, and study the relation between 
this group averaging and the group averaging using the master 
constraint. Finally we prove that under some technical assumptions, the 
two approaches lead to the same result.

In section \ref{conc}, we summarize and conclude.

\section{Group averaging rigging inner product and direct integral 
decomposition }\label{1}

We first consider the master constraint programme. 
Recall that given the self-adjoint master constraint operator $\M$, we 
can formally write down the quantum master constraint equation by 
\be
\M\ \Psi=0\label{MEQ}
\ee
The space of solutions for this equation combined with a certain 
physical inner product is called the physical Hilbert space 
$\ch_{Phys}$. However, the equation Eq.(\ref{MEQ}) is only formal 
because zero is generically contained in the continuous spectrum of 
the master constraint operator, so that the solution state $\Psi$ does 
not 
live in the kinematical Hilbert space anymore. In order to rigorously 
define the space of solutions and to specify the physical inner product, 
we 
should in principle employ the direct integral decomposition (DID) 
\cite{DID} for the master constraint operator $\M$. Whenever  
the master constraint operator $\M$ can be quantized as a self-adjoint
operator, the physical 
Hilbert space $\ch_{Phys}$ is well-defined in principle (modulo measure 
theoretic subtleties which require further physical input but do not 
present mathematicales obstacles).

In \cite{DID}, the programme of direct integral decomposition (DID) is 
introduced in order to rigorously define the physical Hilbert space for 
a general constraint system. It proceeds as the follows: 

\begin{enumerate}

\item Given a kinematical Hilbert space $\mathcal{H}_{Kin}$ and a 
self-adjoint master constraint operator 
$\mathbf{M}=K^{IJ}C^\dagger_IC_J$, we have to first of all split the 
kinematical Hilbert space into three mutually orthogonal sectors 
$\mathcal{H}_{Kin}=\mathcal{H}^{pp}\oplus\mathcal{H}^{ac}\oplus
\mathcal{H}^{cs}$ 
with respect to the three different possible spectral types of the 
master constraint operator $\mathbf{M}$.

\item We make the direct integral decomposition of each $\mathcal{H}^*$, 
$*=pp,ac,cs$ with respect to the spectrum of the master constraint 
operator $\M$ restricted in each sector, i.e.
\begin{eqnarray}
\mathcal{H}^*=\int^\oplus\rmd\mu^*(\l)\ch^{\ast}_\l\nonumber
\end{eqnarray}

\item Finally, we define the physical Hilbert space to be a direct sum 
of 
three fiber Hilbert spaces at $\lambda=0$ with respect to the three 
kinds of 
spectral types, i.e. 
$\mathcal{H}_{Phys}=\mathcal{H}^{pp}_{\lambda=0}
\oplus\mathcal{H}^{ac}_{\lambda=0}\oplus\mathcal{H}^{cs}_{\lambda=0}$
\end{enumerate}
Note that in step 2. we have assumed that all the ambiguities outlined 
in \cite{DID} have been solved by considering some physical criterion 
e.g. the physical Hilbert space should admit sufficiently many 
semiclassical states, 
and it should represent the algebra of Dirac observables as an algebra 
of self-adjoint operators. With this assumption, the procedure of the 
DID 
programme gives a proper definition of the physical Hilbert space for a 
general constraint system. In many models simpler than GR, such a 
programme gives satisfactory results \cite{DID2}.

However, if we want to practically obtain the physical Hilbert space of 
LQG and get detailed knowledge about the structure of this physical 
Hilbert space, then DID is not a suitable procedure. The reason is the 
following: the whole procedure of DID depends on the precise knowledge 
of the 
spectral structure for the master constraint operator. For the
case of LQG or AQG \cite{AQGIV} with a complicated master constraint 
operator $\M$, 
the spectrum of $\M$ is largely unknown so that the DID programme is too 
hard to apply practically. Therefore, for  
practical purposes, we have 
to employ a technique such that the final structure of physical Hilbert 
space 
$\mathcal{H}_{Phys}=\mathcal{H}^{pp}_{\lambda=0}\oplus
\mathcal{H}^{ac}_{\lambda=0}\oplus\mathcal{H}^{cs}_{\lambda=0}$ 
is obtained without much of the knowledge for the spectrum of the master 
constraint operator. Fortunately, we have a single constraint in the 
quantum theory, whose ``gauge transformations'' that it generates form a 
one-parameter group\footnote{These are only gauge transformations 
in the mathematical sense. The Hamiltonian vector field of the classical 
Master 
constraint vanishes on the constraint surface.}.  
Therefore we can employ an alternative, (modified) 
group averaging technique to obtain the physical inner product as 
outlined in the introduction. 

\begin{Definition} \label{GA0}
For each state $\psi$ in a dense subset $\mathcal{D}$ of 
$\mathcal{H}_{Kin}$, a linear functional $\eta_\Omega(\psi)$ in the 
algebraic dual of $\mathcal{D}$ is defined by 
\begin{eqnarray}
\eta_{\Omega}(\psi)[\phi]:=\lim_{\epsilon\rightarrow0}\frac{\int_{\mathbb{R}}\mathrm{d}t\ 
\langle\psi|e^{it(\mathbf{M}-\epsilon)}|\phi\rangle_{Kin}}{\int_{\mathbb{R}}\mathrm{d}t\ 
\langle\Omega|e^{it(\mathbf{M}-\epsilon)}|\Omega\rangle_{Kin}}\nonumber
\end{eqnarray}
$\forall \phi\in\mathcal{D}$ and 
where $\Omega\in\mathcal{H}_{Kin}$ is a once and for all fixed  
reference vector which corresponds to a choice of normalization. 
The inner product on the linear span of the  
$\eta_\Omega(\psi)$ is defined by
$\langle\eta(\psi)|\eta(\phi)\rangle_\Omega:=\eta_{\Omega}(\psi)[\phi]$. 
The resulting Hilbert space is denoted by $\mathcal{H}_\Omega$
\end{Definition}
The reason for taking the limit $\epsilon\to 0$ in this 
definition is in order to establish the connection between the group 
averaging 
Hilbert 
space $\mathcal{H}_\Omega$ and one of the sectors in the physical 
Hilbert 
space as defined via DID above. This will become clear below.

Here we explicitly construct the direct integral decomposition for 
$\mathbf{M}$. We denote by $E(\lambda)$ the projection valued measure 
associated with $\mathbf{M}$, which is a map from the natural Borel 
$\sigma$-algebra on $\mathbb{R}$ into the set of projection operators on 
$\mathcal{H}_{Kin}$. Thus we have a spectral measure for any unit vector 
$\Omega\in\mathcal{H}_{Kin}$ defined by 
\begin{eqnarray}
\mu_\Omega(B)\ =\ \langle\Omega|E(B)|\Omega\rangle_{Kin}\nonumber
\end{eqnarray}
for any measurable set $B$ in $\mathbb{R}$.

Thus the kinematical Hilbert space $\mathcal{H}_{Kin}$ can be decomposed 
into $\mathcal{H}^{pp}\oplus\mathcal{H}^{ac}\oplus\mathcal{H}^{cs}$, 
where $\mathcal{H}^*=\{\Omega\in\mathcal{H}_{Kin}|\ 
\mu_\Omega=\mu_\Omega^*,\ *=pp,ac,cs\ \}$. In each of $\mathcal{H}^*$, 
the projection valued measure of $\mathbf{M}|_{\mathcal{H}^*}$ is 
denoted by $E^*(\lambda)$. Given $\psi_*\in\mathcal{H}^*$ and a smooth 
function with compact support $f\in C^\infty_c(\mathbb{R}^N)$, one can 
construct a $C^\infty$-vector for $\mathbf{M}|_{\mathcal{H}^*}$ by
\begin{eqnarray}
\Omega^{\psi_*}_f:=\int_{\mathbb{R}}\mathrm{d}t\ 
f(t)e^{it\mathbf{M}}\psi_*\nonumber
\end{eqnarray}
and $i\mathbf{M}\ 
\Omega^{\psi_*}_f=-\Omega^{\psi_*}_{\mathrm{d}f/\mathrm{d}t}$. The 
span of these $C^\infty$-vectors as $\psi^*$ and $f$ vary is 
dense in $\mathcal{H}^*$.

Suppose we pick a $C^\infty$-vector $\Omega^*_1$, then we obtain a 
subspace $\mathcal{H}_1^*$ by the closed linear span of the vectors 
$p^*(\mathbf{M})\Omega^*_1$  where $p^*(\mathbf{M})$ 
denotes a polynomial of $\mathbf{M}$. If 
$\mathcal{H}_1^*\neq\mathcal{H}^*$, we can pick another 
$C^\infty$-vector $\Omega^*_2\in\mathcal{H}_1^{*\bot}$ and construct 
another subspace $\mathcal{H}_2^*\subset\mathcal{H}_1^{*\bot}$ in the 
same way. Iterating this procedure, we arrive at an at most countable 
direct sum by the 
separability of $\mathcal{H}^*$
\begin{eqnarray}
\mathcal{H}^*=\oplus_{m=1}^\infty\mathcal{H}_m^*\nonumber
\end{eqnarray}
in which a dense set of vectors can be given in the form 
$\{p^*_m(\mathbf{M})\Omega^*_m\}_{m=1}^\infty$ where each $p^*_m$ is a 
polynomial of $\mathbf{M}$ and each $\Omega^*_m$ is a $C^\infty$-vector 
for $\mathbf{M}$.

For any measurable set $B$ in $\mathbb{R}$, we consider the spectral 
measure 
\begin{eqnarray}
\mu^*_{m}(B)&=&\langle{\Omega}^*_m|\ E^*(B)\ 
|{\Omega}^*_m\rangle^*\nonumber
\end{eqnarray}
If we choose a probability spectral measure $\mu^*=\sum_{m=1}^\infty 
c_m\mu^*_{\Omega_m}$ ($\sum_{m=1}^\infty c_m=1$) with the maxmality 
feature: for any $\psi\in\mathcal{H}^*$ the associated spectral measure 
$\mu^*_\psi(B)\ =\ \langle\psi|E^*(B)|\psi\rangle^*$ is absolutely 
continuous with respect to $\mu^*$ (e.g. if $c_m>0$ for all $m$), we 
have 
\begin{eqnarray}
\mathrm{d}\mu^*_{m}(\lambda)&=&\rho^*_{m}(\lambda)\mathrm{d}\mu^*(\lambda)
\nonumber
\end{eqnarray}
and each $\rho^*_{m}$ is a nonnegative $L^1(\mathbb{R},\mu^*)$ function. 
We will assume that each $\rho^{ac}_m$ has a representative 
which is continuous at $\lambda=0$ 
\cite{DID,DID2}.  

We define the function $N^*: \mathbb{R}\to\mathbb{N}$ by 
$N^*(\lambda)=M$ provided that $\lambda$ lies in precisely $M$ of the 
$S_{\rho^*_{m}}=\{\lambda\in\mathbb{R}|\ \rho^*_{m}(\lambda)>0\}$. We 
also denote by  
$X^*_M$ the pre-image 
$X^*_M=\{\lambda\in\mathbb{R}|N^*(\lambda)=M\}$ of $\{M\}$ under $N^*$. 

For any two vectors $\psi_*=\{p^*_m(\mathbf{M})\Omega^*_m\}_{m}$ and 
$\psi'_*=\{p'^*_m(\mathbf{M})\Omega^*_m\}_{m}$
\begin{eqnarray}
\langle\psi_*|\psi'_*\rangle^*&=&\sum_{m=1}^\infty\langle\Omega^*_m|p^*_m(\mathbf{M})^\dagger 
p'^*_m(\mathbf{M})|\Omega^*_m\rangle^*\nonumber\\
&=&\sum_{m=1}^\infty\int_{\mathbb{R}}\mathrm{d}\mu^*_m(\lambda)\ 
\overline{p^*_m(\lambda)}\ p'^*_m(\lambda)\nonumber\\
&=&\int_{\mathbb{R}}\mathrm{d}\mu^*_m(\lambda)\sum_{m=1}^\infty\rho^*_{m}(\lambda)\ 
\overline{p^*_m(\lambda)}\ p'^*_m(\lambda)\nonumber\\
&=&\sum_{M=1}^\infty\int_{X^*_M}\mathrm{d}\mu^*(\lambda)\sum_{k=1}^{N^*(\lambda)}\rho^*_{m_k(\lambda)}(\lambda)\ 
\overline{p^*_{m_k(\lambda)}(\lambda)}\ 
p'^*_{m_k(\lambda)}(\lambda)\nonumber
\end{eqnarray}
where $\rho^*_{m_k(\lambda)}(\lambda)\neq0$ at $\lambda$. Therefore we 
arrive at a direct integral representation, i.e.
\begin{eqnarray}
\mathcal{H}^*&\simeq&\mathcal{H}^{*,\oplus}_{\mu^*,N^*}\ =\ 
\int^\oplus_{\mathbb{R}}\mathrm{d}\mu^*(\lambda)\ 
\mathcal{H}^*_{\lambda},\nonumber\\
\langle\psi_* |\psi'_* 
\rangle^*&=&\sum_{M=1}^\infty\int_{X^*_M}\mathrm{d}\mu^*(\lambda)\ 
\langle\psi_*(\lambda) |\psi'_*(\lambda) \rangle^*_{\lambda}\label{repM}
\end{eqnarray}
where 
\begin{eqnarray}
\psi_*(\lambda)&=&\sum_{k=1}^{N^*(\lambda)}\sqrt{\rho^*_{m_k(\lambda)}(\lambda)}\ 
p^*_{m_k(\lambda)}(\lambda)e_k(\lambda)\nonumber\\
\langle\psi_*(\lambda) |\psi'_*(\lambda) 
\rangle^*_{\lambda}&=&\sum_{k=1}^{N^*(\lambda)}\rho^*_{m_k(\lambda)}(\lambda)\ 
\overline{p^*_{m_k(\lambda)}(\vec{x})}\ 
p'^*_{m_k(\lambda)}(\lambda)\nonumber
\end{eqnarray}
$\{e_k(\lambda)\}_{k=1}^{N^*(\lambda)}$ is an orthonormal basis in 
$\mathcal{H}^*_{\lambda}\simeq\mathbb{C}^{N^*(\lambda)}$. Then we have 
the following theorem:

\begin{Theorem} \label{GADID}
We suppose zero is not a limit point in $\sigma^{pp}(\mathbf{M})$ and 
that\footnote{The physical interpretation of the continuous singular
spectrum is typically obscure and there exists a wide literature on 
sufficient conditions for its absence \cite{Simon}.}
$\sigma^{cs}(\mathbf{M})=\emptyset$. In addition, if we 
have any one of the following conditions
\begin{enumerate}
\item there exists $\delta>0$ such that each $\mu^{ac}_m$ 
($\mathrm{d}\mu^{ac}_m=\mu^{ac}_m\mathrm{d}\lambda$) is continuous on 
the closed interval $[0,\delta]$.
\item there exists $\delta>0$ such that each $\rho^{ac}_m$ is continuous 
at $\lambda=0$ and is differentiable on the open interval $(0,\delta)$.
\item there exists $\delta>0$ such that $N^{ac}$ is constant on the 
neighborhood $[0,\delta)$.
\end{enumerate}
Then there exists a dense domain $\mathcal{D}$ in $\mathcal{H}_{Kin}$, 
such that for some choice of reference vector $\Omega$ the 
group averaging Hilbert space $\mathcal{H}_\Omega$ is unitarily 
equivalent to the absolutely continuous sector of physical Hilbert space 
$\mathcal{H}^{ac}_{\lambda=0}$.
\end{Theorem}
\textbf{Proof:} First of all, for any two states 
$\psi_{*},\phi_{*}\in\mathcal{H}^*$ ($*=pp,ac$), we consider the 
integral, 
\begin{eqnarray}
&&\int_{\mathbb{R}}\mathrm{d}t\langle\psi_{*}|e^{it(\mathbf{M}-\epsilon)} 
|\phi_{*} \rangle^{*}\nonumber\\
&=&\int_{\mathbb{R}}\mathrm{d}t\int_{\sigma(\mathbf{M})}
\mathrm{d}\mu^{*}(\lambda)\ 
e^{it(\lambda-\epsilon)}\ \langle\psi_{*}(\lambda) |\phi_{*}(\lambda) 
\rangle^{*}_\lambda\nonumber\\
&=&\lim_{g\to0}\int_{\mathbb{R}}\mathrm{d}t\int_{\sigma(\mathbf{M})}
\mathrm{d}\mu^{*}(\lambda)\ 
e^{it(\lambda-\epsilon)-|gt|}\ \langle\psi_{*}(\lambda) 
|\phi_{*}(\lambda) \rangle^{*}_\lambda. \label{manipulation}
\end{eqnarray}
This equation is justified by the Lebesgue monotone convergence theorem 
\cite{Rudin}, because $\{e^{-|gt|}\}_g$ is an monotone increasing family 
for each $t\in\mathbb{R}$ when $g\to0$, and the other part of the 
function in the integrand can be uniquely split into the form 
$u_+(\lambda)-u_-(\lambda)+iv_+(\lambda)-iv_-(\lambda)$ where $u_\pm$ 
and $v_\pm$ are nonnegative measurable functions.

The integrals $\int_{\mathbb{R}}\mathrm{d}t$ and 
$\int_{\sigma(\mathbf{M})}\mathrm{d}\mu^{ac}(\lambda)$ in the above 
equation can be interchanged by Fubini's theorem \cite{Rudin}, since 
the 
integrand $e^{it(\lambda-\epsilon)-|gt|}\ \langle\psi_{*}(\lambda) 
|\phi_{*}(\lambda) \rangle^{*}_\lambda$ is measurable on $\mathbb{R}^2$ 
and 
$\int_{\mathbb{R}}\mathrm{d}t\int_{\sigma(\mathbf{M})}\mathrm{d}\mu^{*}(\lambda)\ 
u_\pm(\lambda)e^{-|gt|}<\infty$ (also for $v_\pm$). Therefore
\begin{eqnarray}
&&\int_{\mathbb{R}}\mathrm{d}t\langle\psi_{*}|e^{it(\mathbf{M}-\epsilon)} 
|\phi_{*} \rangle^{*}\nonumber\\
&=&\lim_{g\to0}\int_{\sigma(\mathbf{M})}\mathrm{d}\mu^{*}(\lambda)\ 
\langle\psi_{*}(\lambda) |\phi_{*}(\lambda) 
\rangle^{*}_\lambda\int_{\mathbb{R}}\mathrm{d}t\ 
e^{it(\lambda-\epsilon)-|gt|}\nonumber\\
&=&\lim_{g\to0}2\int_{\sigma(\mathbf{M})}\mathrm{d}\mu^{*}(\lambda)\ 
\langle\psi_{*}(\lambda) |\phi_{*}(\lambda) 
\rangle^{*}_\lambda\frac{g}{g^2+(\lambda-\epsilon)^2}\nonumber
\end{eqnarray}

We first consider the pure point spectrum $*=pp$. By the assumption that 
zero is not a limit point in $\sigma^{pp}(\mathbf{M})$, for sufficiently 
small $\epsilon$ we have
$\lambda-\epsilon\neq0$ for all $\lambda\in\sigma^{pp}(\mathbf{M})$. 
Then, the 
function $\frac{2g}{g^2+(\lambda-\epsilon)^2}$ is bounded in the limit 
$g\to0$. Therefore the above integral vanishes as one sees by 
applying the Lebesgue 
dominated convergence theorem. Hence for any three states 
$\psi,\phi,\Omega\in\mathcal{H}_{Kin}$
\begin{eqnarray}
\langle\eta(\psi)|\eta(\phi)\rangle_\Omega
&=&\lim_{\epsilon\rightarrow0}\frac{\int_{\mathbb{R}}\mathrm{d}t\ 
\langle\psi|e^{it(\mathbf{M}-\epsilon)}|\phi\rangle^{ac}}{\int_{\mathbb{R}}\mathrm{d}t\ 
\langle\Omega|e^{it(\mathbf{M}-\epsilon)}|\Omega\rangle^{ac}}\nonumber
\end{eqnarray}
so we only need to consider the absolutely continuous spectrum in what 
follows.

Furthermore, we have already seen that 
\begin{eqnarray}
\langle\psi_{ac}|\psi'_{ac}\rangle^{ac}
&=&\sum_{m=1}^\infty\int_{\mathbb{R}}\mathrm{d}\mu^{ac}_m(\lambda)\ 
\overline{p^{ac}_m(\lambda)}\ p'^{ac}_m(\lambda)\nonumber
\end{eqnarray}
for $\psi_{ac}=\sum_{m=1}^\infty p^{ac}_m(\mathbf{M})\Omega^{ac}_m$ and 
$\psi'_{ac}=\sum_{m=1}^\infty p'^{ac}_m(\mathbf{M})\Omega^{ac}_m$ where 
$p^{ac}_m$ and $p'^{ac}_m$ are measurable functions. From this we can 
select a dense domain $\mathcal{S}$ in $\mathcal{H}^{ac}$ by considering 
those 
$\psi_{ac}\in\mathcal{S}$, $\psi_{ac}=\sum_{m=1}^\infty 
p^{ac}_m(\mathbf{M})\Omega^{ac}_m$ with only finitely many of 
the $p^{ac}_m$ 
nonvanishing and such that each $p^{ac}_m\in C^\infty_c(\mathbb{R})$ 
(the set of complex valued functions of compact support).

Choosing in (\ref{manipulation})
$\psi_{ac}=\sum_{m=1}^\infty p^{ac}_m(\mathbf{M})\Omega^{ac}_m$ and 
$\psi'_{ac}=\sum_{m=1}^\infty p'^{ac}_m(\mathbf{M})\Omega^{ac}_m$ in 
$\mathcal{S}$
\begin{eqnarray}
&&\int_{\mathbb{R}}\mathrm{d}t\langle\psi_{ac}|e^{it(\mathbf{M}-\epsilon)} 
|\phi_{ac} \rangle^{ac}\nonumber\\
&=&\int_{\mathbb{R}}\mathrm{d}t\sum_{m=1}^\infty\int_{\mathbb{R}}\mathrm{d}\mu^{ac}_m(\lambda)\ 
e^{it(\lambda-\epsilon)}\ \overline{p^{ac}_m(\lambda)}\ 
p'^{ac}_m(\lambda)\nonumber\\
&=&\sum_{m=1}^\infty\lim_{g\to0}\int_{\mathbb{R}}\mathrm{d}t\int_{\mathbb{R}}\mathrm{d}\mu^{ac}_m(\lambda)\ 
e^{it(\lambda-\epsilon)-|gt|}\ \overline{p^{ac}_m(\lambda)}\ 
p'^{ac}_m(\lambda)\nonumber\\
&=&\sum_{m=1}^\infty\lim_{g\to0}\int_{\mathbb{R}}\mathrm{d}\mu^{ac}_m(\lambda)\ 
\overline{p^{ac}_m(\lambda)}\ 
p'^{ac}_m(\lambda)\frac{2g}{g^2+(\lambda-\epsilon)^2}\nonumber
\end{eqnarray}
Note that the above sum over $m$ is actually a finite sum which is 
why we were allowed to interchange it with the integral.\\
\\
i.\\
Suppose 
condition 1 holds: there exists $\delta>0$ such that each 
$\mu^{ac}_m$ ($\mathrm{d}\mu^{ac}_m=\mu^{ac}_m\mathrm{d}\lambda$) is 
continuous on the closed interval $[0,\delta]$. Then the function 
$\mu^{ac}_m \overline{p^{ac}_m} p'^{ac}_m$ is continuous on the closed 
interval $[0,\delta]$ thus is also bounded on $[0,\delta]$. So if we 
choose $0<\epsilon<\delta$ then
\begin{eqnarray}
\int_{\mathbb{R}}\mathrm{d}t\langle\psi_{ac}|e^{it(\mathbf{M}-\epsilon)} 
|\phi_{ac} \rangle^{ac}
&=&2\pi\sum_{m=1}^\infty\mu^{ac}_m(\epsilon)\ 
\overline{p^{ac}_m(\epsilon)}\ p'^{ac}_m(\epsilon).\nonumber
\end{eqnarray}
Hence for any three states 
$\psi,\phi,\Omega\in\mathcal{D}:=\mathcal{H}^{pp}\oplus\mathcal{S}$ 
($\Omega_{ac}=\sum_{m=1}^\infty f^{ac}_m(\mathbf{M})\Omega^{ac}_m$)
\begin{eqnarray}
\langle\eta(\psi)|\eta(\phi)\rangle_\Omega
&=&\lim_{\epsilon\rightarrow0}\frac{\int_{\mathbb{R}}\mathrm{d}t\ 
\langle\psi_{ac}|e^{it(\mathbf{M}-\epsilon)}|\phi_{ac}\rangle^{ac}}{\int_{\mathbb{R}}\mathrm{d}t\ 
\langle\Omega_{ac}|e^{it(\mathbf{M}-\epsilon)}|\Omega_{ac}\rangle^{ac}}\ 
=\ 
\lim_{\epsilon\rightarrow0}\frac{2\pi\sum_{m=1}^\infty\mu^{ac}_m(\epsilon)\ 
\overline{p^{ac}_m(\epsilon)}\ 
p'^{ac}_m(\epsilon)}{2\pi\sum_{m=1}^\infty\mu^{ac}_m(\epsilon)\ 
\overline{f^{ac}_m(\epsilon)}\ f^{ac}_m(\epsilon)}\nonumber\\
&=&\frac{\sum_{m=1}^\infty\mu^{ac}_m(0)\ \overline{p^{ac}_m(0)}\ 
p'^{ac}_m(0)}{\sum_{m=1}^\infty\mu^{ac}_m(0)\ \overline{f^{ac}_m(0)}\ 
f^{ac}_m(0)}\ =\ 
\frac{\langle\psi_{ac}(0)|\psi'_{ac}(0)\rangle^{ac}_{\lambda=0}}{\langle\Omega_{ac}(0)|\Omega_{ac}(0)\rangle^{ac}_{\lambda=0}}\nonumber
\end{eqnarray}
by using $\mu^{ac}_m(0)=\mu^{ac}(0)\rho^{ac}_m(0)$ as follows from
$d\mu^{ac}_m=\mu^{ac}_m\;d\lambda=\rho^{ac}_m d\mu^{ac}=\rho^{ac} 
\mu^{ac}\;d\lambda$ and $\mu^{ac}(0)>0$ w.l.g.\\
\\
ii.\\
Suppose that condition 2 holds: there exists $\delta>0$ such that 
each 
$\rho^{ac}_m$ is continuous at $\lambda=0$ and is differentiable on the 
open interval $(0,\delta)$. We choose $0<\epsilon<\eta<\delta$ and 
calculate
\begin{eqnarray}
&&\lim_{g\to0}\int_{\mathbb{R}}\mathrm{d}\mu^{ac}(\lambda)\ 
\rho^{ac}_m(\lambda)\ \overline{p^{ac}_m(\lambda)}\ 
p'^{ac}_m(\lambda)\frac{2g}{g^2+(\lambda-\epsilon)^2}\nonumber\\
&=&\lim_{g\to0}\int_0^\eta\mathrm{d}\mu^{ac}(\lambda)\ 
\rho^{ac}_m(\lambda)\ \overline{p^{ac}_m(\lambda)}\ 
p'^{ac}_m(\lambda)\frac{2g}{g^2+(\lambda-\epsilon)^2}\nonumber\\
&=&\lim_{g\to0}\int_0^\eta\mathrm{d}\mu^{ac}(\lambda)\ \frac{ 
\rho^{ac}_m(\lambda)\ \overline{p^{ac}_m(\lambda)}\ 
p'^{ac}_m(\lambda)-\rho^{ac}_m(\epsilon)\ \overline{p^{ac}_m(\epsilon)}\ 
p'^{ac}_m(\epsilon)}{\lambda-\epsilon}\ 
\frac{2g(\lambda-\epsilon)}{g^2+(\lambda-\epsilon)^2}\nonumber\\
&&+\lim_{g\to0}\int_0^\eta\mathrm{d}\mu^{ac}(\lambda)\frac{2g}{g^2+(\lambda-\epsilon)^2}\ 
\rho^{ac}_m(\epsilon)\ \overline{p^{ac}_m(\epsilon)}\ 
p'^{ac}_m(\epsilon) \label{calculation}
\end{eqnarray}
In the second step we have split the integral over $\lambda \in 
\mathbb{R}_+$ into $[0,\eta]$ and $(\eta,\infty)$. The function 
$g/(g^2+(\lambda-\epsilon)^2)$ for $\lambda>\eta$ is bounded from above 
by $g/(g^2+(\eta-\epsilon)^2)$. Therefore the integral restricted 
to $(\eta,\infty)$ is bounded by $2\; 
||\psi_{ac}||\;||\psi'_{ac}||\;g/(g^2+(\eta-\epsilon)^2)$ which 
obviously vanishes as $g\to 0$. Now consider the last line in 
(\ref{calculation}) which consists of two terms. In the integrand of the 
first term, 
$|\frac{2g(\lambda-\epsilon)}{g^2+(\lambda-\epsilon)^2}|\leqslant1$ and 
$\frac{ \rho^{ac}_m(\lambda)\ \overline{p^{ac}_m(\lambda)}\ 
p'^{ac}_m(\lambda)-\rho^{ac}_m(\epsilon)\ \overline{p^{ac}_m(\epsilon)}\ 
p'^{ac}_m(\epsilon)}{\lambda-\epsilon}$ is also bounded on $[0,\eta]$ 
since $\rho^{ac}_m$ is differentiable in at $\lambda=\epsilon$. 
Therefore the integrand in the first term is bounded by a finite 
constant. Thus by Lebesgue dominated convergence theorem we can apply 
the limit $g\to 0$ directly to the integrand. Now the function 
$f(a,b)=2ab/(a^2+b^2)$ for $b\not=0$ has the limit zero for $a\to 0$ and 
$f(a,0)=0$ anyway. Hence the first
term vanishes in the limit. Thus
\begin{eqnarray}
\int_{\mathbb{R}}\mathrm{d}t\langle\psi_{ac}|e^{it(\mathbf{M}-\epsilon)} 
|\phi_{ac} \rangle^{ac}
&=&\lim_{g\to0}\int_0^\eta\mathrm{d}\mu^{ac}(\lambda)\frac{2g}{g^2+(\lambda-\epsilon)^2}\sum_{m=1}^\infty 
\rho^{ac}_m(\epsilon)\ \overline{p^{ac}_m(\epsilon)}\ 
p'^{ac}_m(\epsilon)\nonumber
\end{eqnarray}
So for any three states 
$\psi,\phi,\Omega\in\mathcal{D}=\mathcal{H}^{pp}\oplus\mathcal{S}$ 
we arrive at the same result as above 
(with $\Omega_{ac}=\sum_{m=1}^\infty f^{ac}_m(\mathbf{M})\Omega^{ac}_m$)
\begin{eqnarray}
\langle\eta(\psi)|\eta(\phi)\rangle_\Omega
&=&\lim_{\epsilon\rightarrow0}\frac{\int_{\mathbb{R}}\mathrm{d}t\ 
\langle\psi_{ac}|e^{it(\mathbf{M}-\epsilon)}|\phi_{ac}\rangle^{ac}}{\int_{\mathbb{R}}\mathrm{d}t\ 
\langle\Omega_{ac}|e^{it(\mathbf{M}-\epsilon)}|\Omega_{ac}\rangle^{ac}}
\ =\ \frac{\sum_{m=1}^\infty\rho^{ac}_m(0)\ \overline{p^{ac}_m(0)}\ 
p'^{ac}_m(0)}{\sum_{m=1}^\infty\rho^{ac}_m(0)\ \overline{f^{ac}_m(0)}\ 
f^{ac}_m(0)}\nonumber\\
&=&\frac{\langle\psi_{ac}(0)|\psi'_{ac}(0)\rangle^{ac}_{\lambda=0}}{\langle\Omega_{ac}(0)|\Omega_{ac}(0)\rangle^{ac}_{\lambda=0}}\nonumber
\end{eqnarray}
~\\
iii.\\
Now consider condition 3: there exists $\delta>0$ such that 
$N^{ac}$ is constant on the open interval $[0,\delta)$. In this 
case we need some additional tools:

First, we define a vector space $\mathcal{V}$ which consists of certain 
families smooth complex functions of compact support,
\begin{eqnarray}
\mathcal{V}:=\Big\{\{f_n\}_{n=1}^\infty \big|\ f_n\neq0\ \mathrm{only\ 
for\ a\ finite\ number\ of}\ n,\ f_n\in C^\infty_c(\mathbb{R})\ \forall 
n  \Big\}\nonumber
\end{eqnarray}
where $C^\infty_c(\mathbb{R})$ is the set of smooth 
complex valued function of compact support on $\mathbb{R}$. Then we 
choose an orthonormal 
basis 
for each fiber Hilbert space $\mathcal{H}^{ac}_{\lambda}$. Consider the 
functions $e_n$ with $e_n(\lambda)\in\mathcal{H}^{ac}_{\lambda}$ and 
$e_n(\lambda)=0$ for $n>N^{ac}(\lambda)$ such that $\langle 
e_n(\lambda)|e_m(\lambda)\rangle^{ac}_\lambda=\delta_{n,m}$ for 
$m,n\le N^{ac}(\lambda)$ and zero otherwise. The 
$\{e_n(\lambda)\}_{n=1,...,N^{ac}(\lambda)}$ provide an orthonormal 
basis 
in $\mathcal{H}^{ac}_{\lambda}$. We define a linear map $\imath$ 
from the vector space $\mathcal{V}$ to $\mathcal{H}^{ac}$ by  
\begin{eqnarray}
\imath:\mathcal{V}&\to&\mathcal{H}^{ac}\nonumber\\
\{f_n\}_{n=1}^\infty&\mapsto&\imath(\{f_n\}_{n=1}^\infty)\ :=\ 
\Bigg\{\sum_{n=1}^{N^{ac}(\lambda)}f_n(\lambda)e_n(\lambda)\Bigg\}_\lambda\equiv\{\psi_{ac}(\lambda)\}_\lambda\nonumber
\end{eqnarray} 
where $\psi_{ac}\in\mathcal{H}^{ac}$ since its $\mathcal{H}^{ac}$-norm 
is bounded
\begin{eqnarray}
(||\psi_{ac}||^{ac})^2=\int_{\sigma(\mathbf{M})}\mathrm{d}\mu^{ac}(\lambda)
(||\psi_{ac}(\lambda)||^{ac}_\lambda)^2
=\int_{\sigma(\mathbf{M})}\mathrm{d}\mu^{ac}(\lambda)
\sum_{n=1}^{N^{ac}(\lambda)}|f_n(\lambda)|^2<\infty\nonumber
\end{eqnarray}
by the assumption that $f_n\neq0\ \mathrm{only\ for\ a\ finite\ number\ 
of}\ n$, and $ f_n\in C^\infty_c(\mathbb{R})\ \forall n$. The image of 
this map $\imath(\mathcal{V})$ is denoted by $\mathcal{S}$, so that for 
any two states $\psi_{ac}=\imath(\{f_n\}_{n=1}^\infty),\ 
\phi_{ac}=\imath(\{f'_n\}_{n=1}^\infty)$ in $\mathcal{S}$, their fiber 
inner product 
$\langle\psi_{ac}(\lambda)|\phi_{ac}(\lambda)\rangle^{ac}_\lambda=\sum_{n=1}^{N^{ac}(\lambda)}\bar{f}_n(\lambda)f'_n(\lambda)$ 
is a bounded function of compact support (i.e. its real part and 
imaginary 
part are bounded from above and below). Moreover the assumption that
there exists a neighborhood $[0,\delta)$ on which $N^{ac}$ is a 
constant implies that
$\langle\psi_{ac}(\lambda)|\phi_{ac}(\lambda)\rangle^{ac}_\lambda=\sum_{n=1}^{N^{ac}(\lambda)}\bar{f}_n(\lambda)f'_n(\lambda)$ 
is smooth on $[0,\delta)$ by the finiteness of the families 
$\{f_n\}_{n=1}^\infty$ and $\{f'_n\}_{n=1}^\infty$.

We must show that the subset $\mathcal{S}$ is dense in 
$\mathcal{H}^{ac}$: Suppose there is another state 
$\phi_{ac}\in\mathcal{H}^{ac}$ orthogonal to all the states in 
$\mathcal{S}$, i.e. for any 
$\psi_{ac}=\imath(\{f_n\}_{n=1}^\infty)\in\mathcal{S}$, 
\begin{eqnarray}
0\ =\ \langle\psi |\phi 
\rangle^{ac}&=&\int_{\sigma(\mathbf{M})}\mathrm{d}\mu^{ac}(\lambda)\langle\psi_{ac}(\lambda) 
|\phi_{ac}(\lambda) \rangle^{ac}_\lambda\ =\ 
\int_{\sigma(\mathbf{M})}\mathrm{d}\mu^{ac}(\lambda)\sum_{n=1}^{N^{ac}(\lambda)}\langle\psi_{ac}(\lambda) 
|e_n(\lambda)\rangle^{ac}_\lambda\langle e_n(\lambda)|\phi_{ac}(\lambda) 
\rangle^{ac}_\lambda\nonumber\\
&=&\int_{\sigma(\mathbf{M})}\mathrm{d}\mu^{ac}(\lambda)\sum_{n=1}^{N^{ac}(\lambda)}\bar{f}_n(\lambda)\langle 
e_n(\lambda)|\phi_{ac}(\lambda) \rangle^{ac}_\lambda\ \ \ \ \ \ \ \ \ \ 
\ \ \ \ \forall\ \{f_n\}_{n=1}^\infty\in\mathcal{V}.\nonumber
\end{eqnarray}
For any positive integer $n_0$, we can choose the family 
$\{f_n\}_{n=1}^\infty\in\mathcal{V}$ such that all $f_n$ vanish except 
$f_{n_0}$. Therefore
\begin{eqnarray}
\int_{\sigma(\mathbf{M})}\mathrm{d}\mu^{ac}(\lambda)\bar{f}_{n_0}(\lambda)\langle 
e_{n_0}(\lambda)|\phi_{ac}(\lambda) \rangle^{ac}_\lambda=0\ \ \ \ \ \ \ 
\ \ \ \ \ \ \ \forall\ f_{n_0}\in C_c^{\infty}(\mathbb{R})\nonumber
\end{eqnarray}
Note that the function $\langle e_{n_0}(\lambda)|\phi_{ac}(\lambda) 
\rangle^{ac}_\lambda$ has support $\{\lambda\in\sigma(\mathbf{M})\ |\ 
N^{ac}(\lambda)\ge n_0\}$. Since $C_c^{\infty}(\mathbb{R})$ is dense in 
$L^2(\mathbb{R},\mu^{ac})$, the above result implies that $\langle 
e_{n_0}(\lambda)|\phi_{ac}(\lambda) \rangle^{ac}_\lambda$ vanishes 
$\mu^{ac}$-a.e, which means that $\phi_{ac}=0$ in $\mathcal{H}^{ac}$. So 
we have proved that $\mathcal{S}$ is dense in 
$\mathcal{H}^{ac}$.

For any two states $\psi_{ac},\phi_{ac}\in\mathcal{S}$, we first 
consider the integral for the absolutely continuous sector 
($0<\epsilon<\delta$), 
\begin{eqnarray}
&&\int_{\mathbb{R}}\mathrm{d}t\int_{\sigma(\mathbf{M})}\mathrm{d}\mu^{ac}(\lambda)\ 
e^{it(\lambda-\epsilon)}\ \langle\psi_{ac}(\lambda) |\phi_{ac}(\lambda) 
\rangle^{ac}_\lambda\nonumber\\
&=&\lim_{g\to0}\int_{\sigma(\mathbf{M})}\mathrm{d}\mu^{ac}(\lambda)\ 
\langle\psi_{ac}(\lambda) |\phi_{ac}(\lambda) 
\rangle^{ac}_\lambda\int_{\mathbb{R}}\mathrm{d}t\ 
e^{it(\lambda-\epsilon)-|gt|}\nonumber\\
&=&\lim_{g\to0}2\int_{\sigma(\mathbf{M})}\mathrm{d}\mu^{ac}(\lambda)\ 
\langle\psi_{ac}(\lambda) |\phi_{ac}(\lambda) 
\rangle^{ac}_\lambda\frac{g}{g^2+(\lambda-\epsilon)^2}\nonumber\\
&=&\lim_{g\to0}2\int_{\sigma(\mathbf{M})}\mathrm{d}\mu^{ac}(\lambda)\ 
\Bigg[\langle\psi_{ac}(\lambda) |\phi_{ac}(\lambda) 
\rangle^{ac}_\lambda-\langle\psi_{ac}(\epsilon) |\phi_{ac}(\epsilon) 
\rangle^{ac}_{\epsilon}\Bigg]\frac{g}{g^2+(\lambda-\epsilon)^2}\nonumber\\
&&+2\Bigg[\lim_{g\to0}\int_{\sigma(\mathbf{M})}\mathrm{d}\mu^{ac}(\lambda)\ 
\frac{g}{g^2+(\lambda-\epsilon)^2}\Bigg]\langle\psi_{ac}(\epsilon) 
|\phi_{ac}(\epsilon) \rangle^{ac}_{\epsilon}.\label{ac}
\end{eqnarray}
The first term in the last line of Eq.(\ref{ac}) can be 
computed as follows
\begin{eqnarray}
&&\lim_{g\to0}2\int_{\sigma(\mathbf{M})}\mathrm{d}\lambda\ 
\mu^{ac}(\lambda)\ \Bigg[\langle\psi_{ac}(\lambda) |\phi_{ac}(\lambda) 
\rangle^{ac}_\lambda-\langle\psi_{ac}(\epsilon) |\phi_{ac}(\epsilon) 
\rangle^{ac}_{\epsilon}\Bigg]\frac{g}{g^2+(\lambda-\epsilon)^2}\nonumber\\
&=&\lim_{g\to0}2\int_{\sigma(\mathbf{M})}\mathrm{d}\lambda\ 
\mu^{ac}(\lambda)\ \frac{\langle\psi_{ac}(\lambda) |\phi_{ac}(\lambda) 
\rangle^{ac}_\lambda-\langle\psi_{ac}(\epsilon) |\phi_{ac}(\epsilon) 
\rangle^{ac}_{\epsilon}}{\lambda-\epsilon}\ 
\frac{\frac{\lambda-\epsilon}{g}}{1+\frac{(\lambda-\epsilon)^2}{g^2}}.\label{vanish}
\end{eqnarray}
Here in the integrand, 
$|\frac{(\lambda-\epsilon)/{g}}{1+(\lambda-\epsilon)^2/g^2}|\leqslant\frac{1}{2}$ 
and $\frac{\langle\psi_{ac}(\lambda) |\phi_{ac}(\lambda) 
\rangle^{ac}_\lambda-\langle\psi_{ac}(\epsilon) |\phi_{ac}(\epsilon) 
\rangle^{ac}_{\epsilon}}{\lambda-\epsilon}$ is also bounded since 
$\langle\psi_{ac}(\lambda)|\phi_{ac}(\lambda)\rangle^{ac}_\lambda$ is 
differentiable at $\lambda=\epsilon$. Therefore the above integrand is 
bounded by an integrable function which is a finite constant times 
$\mu^{ac}$ (recall that $\mu^\ast$ is a probability measure). Thus by 
Lebesgue dominated convergence theorem we can apply 
the limit directly to the integrand, which shows that 
Eq.(\ref{vanish}) vanishes in the limit. Therefore, we obtain that
\begin{eqnarray}
\int_{\mathbb{R}}\mathrm{d}t\int_{\sigma(\mathbf{M})}\mathrm{d}\mu^{ac}(\lambda)\ 
e^{it(\lambda-\epsilon)}\ \langle\psi_{ac}(\lambda) |\phi_{ac}(\lambda) 
\rangle^{ac}_\lambda
\ =\ 
2\Bigg[\lim_{g\to0}\int_{\sigma(\mathbf{M})}\mathrm{d}\mu^{ac}(\lambda)\ 
\frac{g}{g^2+(\lambda-\epsilon)^2}\Bigg]\langle\psi_{ac}(\epsilon) 
|\phi_{ac}(\epsilon) \rangle^{ac}_{\epsilon}.\nonumber
\end{eqnarray}

Finally we obtain the same result as above:
\begin{eqnarray}
&&\langle\eta(\psi)|\eta(\phi)\rangle_\Omega\nonumber\\
&=&\lim_{\epsilon\rightarrow0}\frac{\int_{\mathbb{R}}\mathrm{d}t\int_{\sigma(\mathbf{M})}\mathrm{d}\mu^{ac}(\lambda)\ 
e^{it(\lambda-\epsilon)}\ \langle\psi_{ac}(\lambda) |\phi_{ac}(\lambda) 
\rangle^{ac}_\lambda}{\int_{\mathbb{R}}\mathrm{d}t\int_{\sigma(\mathbf{M})}\mathrm{d}\mu^{ac}(\lambda)\ 
e^{it(\lambda-\epsilon)}\ \langle\Omega_{ac}(\lambda) 
|\Omega_{ac}(\lambda) \rangle^{ac}_\lambda}\nonumber\\
&=&\lim_{\epsilon\rightarrow0}\lim_{g\to0}\frac{2\Bigg[\int_{\sigma(\mathbf{M})}\mathrm{d}\mu^{ac}(\lambda)\ 
\frac{g}{g^2+(\lambda-\epsilon)^2}\Bigg]\langle\psi_{ac}(\epsilon) 
|\phi_{ac}(\epsilon) 
\rangle^{ac}_{\epsilon}}{2\Bigg[\int_{\sigma(\mathbf{M})}\mathrm{d}\mu^{ac}(\lambda)\ 
\frac{g}{g^2+(\lambda-\epsilon)^2}\Bigg]\langle\Omega_{ac}(\epsilon) 
|\Omega_{ac}(\epsilon) \rangle^{ac}_{\epsilon}}\nonumber\\
&=&\frac{\langle\psi_{ac}(0) |\phi_{ac}(0) 
\rangle^{ac}_{\lambda=0}}{\langle\Omega_{ac}(0) |\Omega_{ac}(0) 
\rangle^{ac}_{\lambda=0}}.\nonumber
\end{eqnarray}

Finally, notice that for any state $\psi_{ac}\in\mathcal{S}$, 
$\psi_{ac}(0)$ is a finite linear span of the $e_n(0)$. The 
linear span of such $\psi_{ac}(0)$ is dense in the Hilbert space 
$\mathcal{H}^{ac}_{\lambda=0}$. Thus we obtain an isometric or conformal 
bijection between $\mathcal{H}^{ac}_{\lambda=0}$ and 
$\mathcal{H}_{\Omega}$ depending on the choice of $\Omega$. Thus for 
suitable $\Omega$ these two Hilbert 
spaces are unitarily equivalent.\\
$\Box$

Now we can see that the reason of taking the limit $\epsilon\to0$ in 
Definition \ref{GA0} is to make the desired connection between the group 
averaging Hilbert space $\mathcal{H}_\Omega$ and the absolutely 
continuous 
sector $\mathcal{H}^{ac}_{\lambda=0}$ in the physical Hilbert space. For 
the 
pure point sector $\mathcal{H}^{pp}_{\lambda=0}$, one should rather 
solve the eigenvalue equation Eq.(\ref{MEQ}) in the kinematical Hilbert 
space 
$\ch_{Kin}$. For the case of LQG, many eigenstates in $\ch_{Kin}$ have 
been found, which correspond to a degenerate geometry, e.g. the 
spin-networks with valence less than 4.
 
It is remarkable that all the physical models  
gravity tested in \cite{DID2} satisfy all the assumptions in Theorem 
\ref{GADID}. This means that the group averaging technique in Definition 
\ref{GA0} gives correct physical Hilbert space (the absolutely 
continuous 
sector) for all those physical models.

\section{The consistency between the group averaging approaches with 
Abelianized constraints and master constraint}\label{2}

\subsection{A finite number of Abelianized constraints }

Now we consider the Dirac quantization for the given system. Suppose we 
have a gauge system with a finite collection of irreducible first class 
constraints $C_I$ ($I=1,2,\cdots,N$, $N$ is finite), then one can 
always locally (in phase space)
abelianize these constraints to obtain 
$\tilde{C}_I=R_{IJ}C_J$, such that $\{\tilde{C}_I,\tilde{C}_J\}=0$ 
\cite{HT}. If we quantize these abelianized constraints as self-adjoint 
operators with $[\tilde{C}_I,\tilde{C}_J]=0$ on $\mathcal{H}_{Kin}$, a 
group averaging approach can be defined due to the Abelian Lie algebra 
structure of the constraint algebra. For each state $\psi$ in a dense 
subset $\mathcal{D}$ of $\mathcal{H}_{Kin}$, a linear functional 
$\eta_\Omega(\psi)$ in the algebraic dual of $\mathcal{D}$ can be 
defined 
such that $\forall \phi\in\mathcal{D}$
\begin{eqnarray}
\eta_{\Omega}(\psi)[\phi]:=\lim_{\epsilon_I\rightarrow0}\frac{\int_{\mathbb{R}}\prod_{I=1}^N\mathrm{d}t_I\ 
\langle\psi|\prod_{I=1}^N 
e^{it_I(\tilde{C}_I-\epsilon_I)}|\phi\rangle_{Kin}}{\int_{\mathbb{R}}\prod_{I=1}^N\mathrm{d}t_I\ 
\langle\Omega|\prod_{I=1}^N 
e^{it_I(\tilde{C}_I-\epsilon_I)}|\Omega\rangle_{Kin}}\nonumber
\end{eqnarray}
where $\Omega\in\mathcal{H}_{Kin}$ is a reference vector. Therefore we 
can define the group averaging inner product on the linear span of 
$\eta_\Omega(\psi)$ via 
$\langle\eta(\psi)|\eta(\phi)\rangle_\Omega:=\eta_{\Omega}(\psi)[\phi]$. 
The resulting Hilbert space is denoted by $\mathcal{H}_\Omega$ 

On the other hand, one can also define a single master constraint 
operator $\mathbf{M}:=K_{IJ}\tilde{C}_I\tilde{C}_J$ where $K_{IJ}$ is a 
positive definite c-number matrix. Therefore a group averaging approach 
can also be defined for the master constraint: For each state $\psi$ in 
the same dense subset $\mathcal{D}$ of $\mathcal{H}_{Kin}$, a linear 
functional $\tilde{\eta}_\Omega(\psi)$ in the algebra dual of 
$\mathcal{D}$ can be defined such that 
$\forall \phi\in\mathcal{D}$
\begin{eqnarray}
\tilde{\eta}_{\Omega}(\psi)[\phi]:=\lim_{\epsilon\rightarrow0}\frac{\int_{\mathbb{R}}\mathrm{d}t\ 
\langle\psi|e^{it(\mathbf{M}-\epsilon)}|\phi\rangle_{Kin}}{\int_{\mathbb{R}}\mathrm{d}t\ 
\langle\Omega|e^{it(\mathbf{M}-\epsilon)}|\Omega\rangle_{Kin}}\nonumber
\end{eqnarray}
where $\Omega\in\mathcal{H}_{Kin}$ is the reference vector. Therefore we 
can define another group averaging inner product on the linear space of 
$\tilde{\eta}_\Omega(\psi)$ via 
$\langle\tilde{\eta}(\psi)|\tilde{\eta}(\phi)\rangle_\Omega:=\tilde{\eta}_{\Omega}(\psi)[\phi]$. 
The resulting Hilbert space is denoted by $\tilde{\mathcal{H}}_\Omega$. 
It is expected that there is consistency between these two 
approaches, since both maps should ``project'' onto the same 
(generalised) kernel. This is what we will establish in
what follows.

As a preparation, we construct the direct integral decomposition with 
respect to the constraints $\tilde{C}_I$. Given this Abelian constraint 
operator algebra, each of these self-adjoint constraints $\tilde{C}_I$ 
($I=1,\cdots,N$) is associated with a projection valued measure $E_I$, 
and $[E_I,E_J]=0$ by commutativity. Then one can define a new 
projection value measure $E=\prod_{I=1}^N\;E_I$, which is a map from the 
natural Borel $\sigma$-algebra on $\mathbb{R}^N$ into the set of 
projection operators on $\mathcal{H}_{Kin}$. Thus we have a spectral 
measure for any unit vector $\Omega\in\mathcal{H}_{Kin}$ defined by 
\begin{eqnarray}
\mu_\Omega(B)\ =\ \langle\Omega|E(B)|\Omega\rangle_{Kin}\nonumber
\end{eqnarray}
for any measurable set $B$ in $\mathbb{R}^N$.

Thus the kinematical Hilbert space $\mathcal{H}_{Kin}$ can be decomposed 
into $\mathcal{H}^{pp}\oplus\mathcal{H}^{ac}\oplus\mathcal{H}^{cs}$, 
where $\mathcal{H}^*=\{\Omega\in\mathcal{H}_{Kin}|\ 
\mu_\Omega=\mu_\Omega^*,\ *=pp,ac,cs\ \}$. In each of $\mathcal{H}^*$, 
the projection valued measure of $\{\tilde{C}_I|_{\mathcal{H}^*}\}_I$ is 
denoted by $E^*(\vec{x})$. Given $\psi_*\in\mathcal{H}^*$ and a smooth 
function with compact support $f\in C^\infty_c(\mathbb{R}^N)$, one can 
construct a $C^\infty$-vector for $\{\tilde{C}_I|_{\mathcal{H}^*}\}_I$ 
by
\begin{eqnarray}
\Omega^{\psi_*}_f:=\int_{\mathbb{R}^N}\mathrm{d}^Nt\ 
f(\vec{t})\prod^N_{I=1}e^{it_I\tilde{C}_I}\psi_*\nonumber
\end{eqnarray}
and $i\tilde{C}_I\Omega^{\psi_*}_f=-\Omega^{\psi_*}_{\partial_If}$. 
Moreover the span of this kind of $C^\infty$-vectors as $\psi^*$ and $f$ 
vary is dense in $\mathcal{H}^*$.

Suppose we pick a $C^\infty$-vector $\Omega_1$, then we obtain a 
subspace $\mathcal{H}_1^*$ by the linear span of 
$q(\{\tilde{C}_I\})\Omega_1$ and completion, where $q(\{\tilde{C}_I\})$ 
denotes a polynomial of $\tilde{C}_I$. If 
$\mathcal{H}_1^*\neq\mathcal{H}^*$, we can pick up another 
$C^\infty$-vector $\Omega_2\in\mathcal{H}_1^{*\bot}$ and construct 
another subspace $\mathcal{H}_2^*\subset\mathcal{H}_1^{*\bot}$ in the 
same way. Iterating this procedure, we arrive at an at most countably 
infinite direct sum 
by the 
separability of $\mathcal{H}^*$
\begin{eqnarray}
\mathcal{H}^*=\oplus_{m=1}^\infty\mathcal{H}_m^*\nonumber
\end{eqnarray}
in which a dense set of vectors can be given in the form 
$\{q_m(\{\tilde{C}_I\})\Omega_m\}_{m=1}^\infty$ where each $q_m$ is a 
polynomial of $\tilde{C}_I$ and each $\Omega_m$ is a $C^\infty$-vectors 
for $\{\tilde{C}_I\}$.

For any measurable set $B$ in $\mathbb{R}^N$, we consider the spectral 
measure 
\begin{eqnarray}
\mu^*_{{\Omega}_m}(B)&=&\langle{\Omega}_m|\ E^*(B)\ 
|{\Omega}_m\rangle^*\nonumber
\end{eqnarray}
If we choose a probability spectral measure $\mu^*=\sum_{m=1}^\infty 
c_m\mu^*_{\Omega_m}$ ($\sum_{m=1}^\infty c_m=1$) with the maxmality 
feature: for any $\psi\in\mathcal{H}^*$ the associated spectral measure 
$\mu^*_\psi(B)\ =\ \langle\psi|E^*(B)|\psi\rangle^*$ is absolutely 
continuous with respect to $\mu^*$, we have 
\begin{eqnarray}
\mathrm{d}\mu^*_{\Omega_m}(\vec{x})&=&\rho^*_{\Omega_m}(\vec{x})\mathrm{d}\mu^*(\vec{x})\nonumber
\end{eqnarray}

We define the function $N^*: \mathbb{R}^N\to\mathbb{N}$ by 
$N^*(\vec{x})=M$ provided that $\vec{x}$ lies in precisely $M$ of the 
$S_{\rho^*_{\Omega_m}}=\{\vec{x}|\ \rho^*_{\Omega_m}(\vec{x})>0\}$. Here 
$X^*_M$ denotes its pre-image 
$X^*_M=\{\vec{x}\in\mathbb{R}^N|N^*(\vec{x})=M\}$ of $\{M\}$. 

For any two vectors $\psi_*=\{q^m(\{\tilde{C}_I\})\Omega_m\}_{m}$ and 
$\psi'_*=\{q'^m(\{\tilde{C}_I\})\Omega_m\}_{m}$
\begin{eqnarray}
\langle\psi_*|\psi'_*\rangle^*&=&\sum_{m=1}^\infty\langle\Omega_m|q^m(\{\tilde{C}_I\})^\dagger 
q'^m(\{\tilde{C}_I\})|\Omega_m\rangle^*\ =\ 
\sum_{m=1}^\infty\int_{\mathbb{R}^N}\mathrm{d}\mu^*_{\Omega_m}(\vec{x})\ 
\overline{q^m(\vec{x})}\ q'^m(\vec{x})\nonumber\\
&=&\int_{\mathbb{R}^N}\mathrm{d}\mu^*(\vec{x})\sum_{m=1}^\infty\rho^*_{\Omega_m}(\vec{x})\ 
\overline{q^m(\vec{x})}\ q'^m(\vec{x})\nonumber\\
&=&\sum_{M=1}^\infty\int_{X^*_M}\mathrm{d}\mu^*(\vec{x})\sum_{k=1}^{N^*(\vec{x})}\rho^*_{\Omega_{m_k(\vec{x})}}(\vec{x})\ 
\overline{q^{m_k(\vec{x})}(\vec{x})}\ 
q'^{m_k(\vec{x})}(\vec{x})\nonumber
\end{eqnarray}
where $\rho^*_{\Omega_{m_k(\vec{x})}}(\vec{x})\neq0$ at $\vec{x}$. 
Therefore we arrive at a direct integral representation, i.e.
\begin{eqnarray}
\mathcal{H}^*&\simeq&\mathcal{H}^{*,\oplus}_{\mu^*,N^*}\ =\ 
\int^\oplus_{\mathbb{R}^N}\mathrm{d}\mu^*(\vec{x})\ 
\mathcal{H}^*_{\vec{x}},\nonumber\\
\langle\psi_* |\psi'_* 
\rangle^*&=&\sum_{M=1}^\infty\int_{X^*_M}\mathrm{d}\mu^*(\vec{x})\ 
\langle\psi_*(\vec{x}) |\psi'_*(\vec{x}) \rangle^*_{\vec{x}}\label{rep1}
\end{eqnarray}
where 
\begin{eqnarray}
\psi_*(\vec{x})&=&\sum_{k=1}^{N^*(\vec{x})}\sqrt{\rho^*_{\Omega_{m_k(\vec{x})}}(\vec{x})}\ 
q^{m_k(\vec{x})}(\vec{x})e_k(\vec{x})\nonumber\\
\langle\psi_*(\vec{x}) |\psi'_*(\vec{x}) 
\rangle^*_{\vec{x}}&=&\sum_{k=1}^{N^*(\vec{x})}\rho^*_{\Omega_{m_k(\vec{x})}}(\vec{x})\ 
\overline{q^{m_k(\vec{x})}(\vec{x})}\ 
q'^{m_k(\vec{x})}(\vec{x})\nonumber
\end{eqnarray}
$\{e_k(\vec{x})\}_{k=1}^{N^*(\vec{x})}$ is an orthonormal basis in 
$\mathcal{H}^*_{\vec{x}}\simeq\mathbb{C}^{N^*(\vec{x})}$.

We are now in the position to prove a result about the relation between 
the  
two group 
averaging approaches (we denote by $\Sigma^*\subset\mathbb{R}^N$ the 
$*$-spectrum of the algebra $\{\tilde{C}_I\}_I$):
\begin{Theorem} \label{FA}
We suppose $\sigma^{cs}(\tilde{C}_I)=\emptyset$, $\vec{x}=0$ is not 
contained in $\Sigma^{cs}$ and is not a limit point in any 
$\sigma^{pp}(\tilde{C}_I)$. We also assume that there exists a 
neighborhood $\mathcal{N}_0$ of $\vec{x}=0$ such that each 
$\rho^{ac}_{\Omega_m}$ is continuous at $\vec{x}=0$ and is 
differentiable on $\mathcal{N}_0-\{\vec{x}=0\}$. With these assumptions, 
the group averaging Hilbert spaces of these two approaches, 
${\mathcal{H}}_\Omega$ and $\tilde{\mathcal{H}}_\Omega$, are unitarily 
equivalent with each other.
\end{Theorem}
\textbf{Proof:} There exists a dense domain 
$\mathcal{D}\subset\mathcal{H}_{Kin}$, such that 
$\mathcal{D}=\mathcal{H}^{pp}\oplus\mathcal{S}\oplus\mathcal{H}^{cs}$ 
and the dense domain $\mathcal{S}$ in $\mathcal{H}^{ac}$ consisting of 
the 
collection 
of all $\psi_{ac}=\sum_{m=1}^\infty q^m(\{\tilde{C}_I\})\Omega_m$ with 
only finitely many $q_m$ nonvanishing and each $q_m$ is a polynomial of 
$\tilde{C}_I$.

For any two vectors $\psi_{ac}=\{q^m(\{\tilde{C}_I\})\Omega_m\}_{m}$ and 
$\psi'_{ac}=\{q'^m(\{\tilde{C}_I\})\Omega_m\}_{m}$ in $\mathcal{S}$
\begin{eqnarray}
&&\int_{\mathbb{R}}\prod_{I=1}^N\mathrm{d}t_I\langle\psi_{ac}|\prod_{I=1}^N 
e^{it_I(\tilde{C}_I-\epsilon_I)}|\psi'_{ac}\rangle^{ac}\nonumber\\
&=&\int_{\mathbb{R}}\prod_{I=1}^N\mathrm{d}t_I\sum_{m=1}^\infty\langle\Omega_m|q^m(\{\tilde{C}_I\})^\dagger\prod_{I=1}^N 
e^{it_I(\tilde{C}_I-\epsilon_I)} 
q'^m(\{\tilde{C}_I\})|\Omega_m\rangle^{ac}\nonumber\\
&=&\int_{\mathbb{R}}\prod_{I=1}^N\mathrm{d}t_I\sum_{m=1}^\infty\int_{\mathbb{R}^N}\mathrm{d}\mu^{ac}_{\Omega_m}(\vec{x})\ 
\prod_{I=1}^N e^{it_I(x_I-\epsilon_I)}\overline{q^m(\vec{x})}\ 
q'^m(\vec{x})\nonumber
\end{eqnarray}
Note that we can freely interchange the sum over $m$ and the integral 
since 
only finite number of terms contribute to the sum. Then as in the 
previous section, we add a 
convergence factor and interchange the integrals
\begin{eqnarray}
&&\int_{\mathbb{R}}\prod_{I=1}^N\mathrm{d}t_I\langle\psi_{ac}|\prod_{I=1}^N 
e^{it_I(\tilde{C}_I-\epsilon_I)}|\psi'_{ac}\rangle^{ac}\nonumber\\
&=&\sum_{m=1}^\infty\lim_{g_I\to0}\int_{\mathbb{R}}\prod_{I=2}^N\mathrm{d}t_I\int_{\mathbb{R}}\mathrm{d}t_1\int_{\mathbb{R}^N}\mathrm{d}\mu^{ac}(\vec{x})\ 
\rho^{ac}_{\Omega_m}(\vec{x})\ \prod_{I=1}^N 
e^{it_I(x_I-\epsilon_I)-|g_It_I|}\ \overline{q^m(\vec{x})}\ 
q'^m(\vec{x})\nonumber\\
&=&\sum_{m=1}^\infty\lim_{g_I\to0}\int_{\mathbb{R}}\prod_{I=3}^N\mathrm{d}t_I\int_{\mathbb{R}}\mathrm{d}t_2\int_{\mathbb{R}^N}\mathrm{d}\mu^{ac}(\vec{x})\ 
\prod_{I=2}^N e^{it_I(x_I-\epsilon_I)-|g_It_I|}\ 
\frac{2g_1}{g_1^2+(x_1-\epsilon_1)^2} \ \rho^{ac}_{\Omega_m}(\vec{x})\ 
\overline{q^m(\vec{x})}\ q'^m(\vec{x})\nonumber\\
&=&\sum_{m=1}^\infty\lim_{g_I\to0}\ \cdots\cdots\nonumber\\
&=&\sum_{m=1}^\infty\lim_{g_I\to0}\int_{\mathbb{R}^N}\mathrm{d}\mu^{ac}(\vec{x})\ 
\prod_{I=1}^N \frac{2g_I}{g_I^2+(x_I-\epsilon_I)^2} \ 
\rho^{ac}_{\Omega_m}(\vec{x})\ \overline{q^m(\vec{x})}\ 
q'^m(\vec{x})\nonumber\\
&=&\sum_{m=1}^\infty\lim_{g_I\to0}\lim_{g_1\to0}\int_{\mathbb{R}^N}\mathrm{d}\mu^{ac}(\vec{x})\ 
\prod_{I=2}^N \frac{2g_I}{g_I^2+(x_I-\epsilon_I)^2}\ 
\frac{2g_1(x_1-\epsilon_1)}{g_1^2+(x_1-\epsilon_1)^2}\nonumber\\
&&\times\frac{ \rho^{ac}_{\Omega_m}(x_1,x_2...,x_N)\ 
\overline{q^m(x_1,x_2...,x_N)}\ 
q'^m(x_1,x_2...,x_N)-\rho^{ac}_{\Omega_m}(\epsilon_1,x_2...,x_N)\ 
\overline{q^m(\epsilon_1,x_2...,x_N)}\ 
q'^m(\epsilon_1,x_2...,x_N)}{x_1-\epsilon_1}\nonumber\\
&&+\sum_{m=1}^\infty\lim_{g_I\to0}\int_{\mathbb{R}^N}\mathrm{d}\mu^{ac}(\vec{x})\ 
\prod_{I=1}^N \frac{2g_I}{g_I^2+(x_I-\epsilon_I)^2}\ 
\rho^{ac}_{\Omega_m}(\epsilon_1,x_2...,x_N)\ 
\overline{q^m(\epsilon_1,x_2...,x_N)}\ 
q'^m(\epsilon_1,x_2...,x_N)\nonumber
\end{eqnarray}
Note that here we choose $\vec{\epsilon}$ contained in a closed N-cube 
$\times_{I=1}^N[-\delta_I,\delta_I]\in\mathcal{N}_0$. Since all 
$\rho^{ac}_{\Omega_m}$, $q^m$ and $q'^m$ are differentiable at 
$\vec{\epsilon}$, the first term in the above result vanishes by the
already familiar reasoning. 
Then we can iterate the same procedure for 
$x_2,...,x_N$ and obtain
\begin{eqnarray}
&&\int_{\mathbb{R}}\prod_{I=1}^N\mathrm{d}t_I\langle\psi_{ac}|\prod_{I=1}^N 
e^{it_I(\tilde{C}_I-\epsilon_I)}|\psi'_{ac}\rangle^{ac}\nonumber\\
&=&\lim_{g_I\to0}\int_{\mathbb{R}^N}\mathrm{d}\mu^{ac}(\vec{x})\ 
\prod_{I=1}^N \frac{2g_I}{g_I^2+(x_I-\epsilon_I)^2}\ 
\sum_{m=1}^\infty\rho^{ac}_{\Omega_m}(\vec{\epsilon})\ 
\overline{q^m(\vec{\epsilon})}\ q'^m(\vec{\epsilon})\nonumber
\end{eqnarray}

Now we consider the pure point sector and continuous singular sector, 
respectively. 
Note 
that since $\sigma^{cs}(\tilde{C}_I)=\emptyset$, for any point 
$(x_1,...,x_N)\in\mathbb{R}^N$ in the continuous singular spectrum 
$\Sigma^{cs}$, there must be at least one $x_I$ taking values in 
$\sigma^{pp}(\tilde{C}_I)$ but not all of them. So $\mathcal{N}_0$ can 
be chosen such that 
$\mathcal{N}_0\cap\Sigma^{*}=\emptyset$, $*=pp,cs$, by the assumption 
that $\vec{x}=0$ is not contained in $\Sigma^{cs}$ and is not a limit 
point in any $\sigma^{pp}(\tilde{C}_I)$. Thus for any two states 
$\psi_{*},\phi_{*}\in\mathcal{H}^*$, $*=pp,cs$ 
\begin{eqnarray}
&&\int_{\mathbb{R}}\prod_{I=1}^N\mathrm{d}t_I\int_{\Sigma^*}\mathrm{d}\mu^{*}(\vec{x})\prod_{I=1}^N 
e^{it_I(x_I-\epsilon_I)} 
\langle\psi_{*}(\vec{x})|\phi_{*}(\vec{x})\rangle^{*}_{\vec{x}}\nonumber\\
&=&\lim_{g_I\to0}\int_{\mathbb{R}}\prod_{I=1}^N\mathrm{d}t_I\int_{\Sigma^*}\mathrm{d}\mu^{*}(\vec{x})\prod_{I=1}^N 
e^{it_I(x_I-\epsilon_I)-|g_It_I|} 
\langle\psi_{*}(\vec{x})|\phi_{*}(\vec{x})\rangle^{*}_{\vec{x}}\nonumber\\
&=&\lim_{g_I\to0}\int_{\Sigma^*}\mathrm{d}\mu^{*}(\vec{x})\prod_{I=1}^N 
\int_{\mathbb{R}}\mathrm{d}t_I\ e^{it_I(x_I-\epsilon_I)-|g_It_I|} 
\langle\psi_{*}(\vec{x})|\phi_{*}(\vec{x})\rangle^{*}_{\vec{x}}\nonumber\\
&=&\lim_{g_I\to0}\int_{\Sigma^*}\mathrm{d}\mu^{*}(\vec{x})\prod_{I=1}^N 
\frac{2g_I}{g_I^2+(x_I-\epsilon_I)^2} 
\langle\psi_{*}(\vec{x})|\phi_{*}(\vec{x})\rangle^{*}_{\vec{x}}\nonumber\\
&=&\lim_{g_I\to0}\int_{\Sigma^*}\mathrm{d}\mu^{*}(\vec{x})\prod_{I=1}^N 
\frac{2g_I}{g_I^2+(x_I-\epsilon_I)^2} 
\langle\psi_{*}(\vec{x})|\phi_{*}(\vec{x})\rangle^{*}_{\vec{x}}\ 
.\nonumber
\end{eqnarray}
Since for sufficiently small $\epsilon$ we have $x_I-\epsilon_I\neq0$ 
for all $\vec{x}\in\Sigma^*$, the function 
$\frac{2g_I}{g_I^2+(x_I-\epsilon_I)^2}$ is bounded in the limit 
$g_I\to0$. Therefore the above integrals vanish in the limit by an 
appeal to Lebesgue dominated convergence theorem.

With the above results, we obtain the following. For any three states 
$\psi,\phi,\Omega\in\mathcal{D}$ 
($\Omega_{ac}=\{f^m(\{\tilde{C}_I\})\Omega_m\}_{m}$),
\begin{eqnarray}
&&\langle\eta(\psi)|\eta(\phi)\rangle_\Omega\nonumber\\
&=&\lim_{\epsilon_I\rightarrow0}\frac{\int_{\mathbb{R}}\prod_I\mathrm{d}t_I\ 
\langle\psi|\prod_I 
e^{it_I(\tilde{C}_I-\epsilon_I)}|\phi\rangle_{Kin}}{\int_{\mathbb{R}}\prod_I\mathrm{d}t_I\ 
\langle\Omega|\prod_I 
e^{it_I(\tilde{C}_I-\epsilon_I)}|\Omega\rangle_{Kin}}\nonumber\\
&=&\lim_{\epsilon_I\rightarrow0}\frac{\sum_{*=pp,ac,cs}\int_{\mathbb{R}}\prod_I\mathrm{d}t_I\int_{\Sigma^*}\mathrm{d}\mu^{*}(\vec{x})\prod_I 
e^{it_I(x_I-\epsilon_I)} 
\langle\psi_{*}(\vec{x})|\phi_{*}(\vec{x})\rangle^{*}_{\vec{x}}}{\sum_{*=pp,ac,cs}\int_{\mathbb{R}}\prod_I\mathrm{d}t_I\int_{\Sigma^*}\mathrm{d}\mu^{*}(\vec{x})\prod_I 
e^{it_I(x_I-\epsilon_I)} 
\langle\Omega_{*}(\vec{x})|\Omega_{*}(\vec{x})\rangle^{*}_{\vec{x}}}\nonumber\\
&=&\lim_{\epsilon_I\rightarrow0}\lim_{g_I\to0}\frac{\Big[\int_{\Sigma^{ac}}\mathrm{d}\mu^{ac}(\vec{x})\prod_{I=1}^N 
\frac{2g_I}{g_I^2+(x_I-\epsilon_I)^2}\Big]\sum_{m=1}^\infty\rho^{ac}_{\Omega_m}(\vec{\epsilon})\ 
\overline{q^m(\vec{\epsilon})}\ 
q'^m(\vec{\epsilon})}{\Big[\int_{\Sigma^{ac}}\mathrm{d}\mu^{ac}(\vec{x})\prod_{I=1}^N 
\frac{2g_I}{g_I^2+(x_I-\epsilon_I)^2}\Big]\sum_{m=1}^\infty\rho^{ac}_{\Omega_m}(\vec{\epsilon})\ 
\overline{f^m(\vec{\epsilon})}\ f^m(\vec{\epsilon})}\nonumber\\
&=&\frac{\langle\psi_{ac}(0)|\phi_{ac}(0)\rangle^{ac}_{\vec{x}=0}}{\langle\Omega_{ac}(0)|\Omega_{ac}(0)\rangle^{ac}_{\vec{x}=0}}\ 
.\nonumber
\end{eqnarray}
Therefore we have obtained an isomorphism from $\mathcal{H}_\Omega$ to 
the
fiber Hilbert space $\mathcal{H}^{ac}_{\vec{x}=0}$ in the absolutely 
continuous sector for a certain choice of the reference vector 
$\Omega$.\\
\\
We now compare this with the group averaging for the master constraint 
$\mathbf{M}$. As before, for any two states 
$\psi_{*},\phi_{*}\in\mathcal{H}^*$, $*=pp,ac,cs$, we compute the 
integral 
\begin{eqnarray}
&&\int_{\mathbb{R}}\mathrm{d}t\int_{\Sigma^*}\mathrm{d}\mu^{*}(\vec{x})\ 
e^{it(K_{IJ}x_Ix_J-\epsilon)} 
\langle\psi_{*}(\vec{x})|\phi_{*}(\vec{x})\rangle^{*}_{\vec{x}}\nonumber\\
&=&\lim_{g\to0}\int_{\mathbb{R}}\mathrm{d}t\int_{\Sigma^*}\mathrm{d}\mu^{*}(\vec{x})\ 
e^{it(K_{IJ}x_Ix_J-\epsilon)-|gt|} 
\langle\psi_{*}(\vec{x})|\phi_{*}(\vec{x})\rangle^{*}_{\vec{x}}\nonumber\\
&=&\lim_{g\to0}\int_{\Sigma^*}\mathrm{d}\mu^{*}(\vec{x})\int_{\mathbb{R}}\mathrm{d}t\ 
e^{it(K_{IJ}x_Ix_J-\epsilon)-|gt|} 
\langle\psi_{*}(\vec{x})|\phi_{*}(\vec{x})\rangle^{*}_{\vec{x}}\nonumber\\
&=&\lim_{g\to0}\int_{\Sigma^*}\mathrm{d}\mu^{*}(\vec{x}) 
\frac{2g}{g^2+(K_{IJ}x_Ix_J-\epsilon)^2} 
\langle\psi_{*}(\vec{x})|\phi_{*}(\vec{x})\rangle^{*}_{\vec{x}}\nonumber
\end{eqnarray}
Here we assume that the sphere\footnote{Of course we assume that the 
matrix 
$K$ is not operator valued but just a positive real valued matrix.} 
$S_\epsilon$ 
defined by 
$K_{IJ}x_Ix_J=\epsilon$ is contained in $\mathcal{N}_0$. Since 
$\mathcal{N}_0$ can be chosen such that 
$\mathcal{N}_0\cap\Sigma^{*}=\emptyset$, $*=pp,cs$, the integrals for 
both pure point sector and continuous singular sector vanish in the 
limit for  
the same reason as before. Therefore,
\begin{eqnarray}
\langle\tilde{\eta}(\psi)|\tilde{\eta}(\phi)\rangle_\Omega&=&\lim_{\epsilon\rightarrow0}\frac{\int_{\mathbb{R}}\mathrm{d}t\ 
\langle\psi_{ac}|e^{it(\mathbf{M}-\epsilon)}|\phi_{ac}\rangle^{ac}}{\int_{\mathbb{R}}\mathrm{d}t\ 
\langle\Omega_{ac}|e^{it(\mathbf{M}-\epsilon)}|\Omega_{ac}\rangle^{ac}}\nonumber
\end{eqnarray}
where we have now reduced the problem to a single sector 
$\mathcal{H}^{ac}$ 
on which $\mathbf{M}$ only has absolutely continuous spectrum.

Given two vectors $\psi_{ac}$ and $\psi'_{ac}$ in $\mathcal{S}$ which 
can be written as $\psi_{ac}=\{q^m(\{\tilde{C}_I\})\Omega_m\}_{m}$ and 
$\psi'_{ac}=\{q'^m(\{\tilde{C}_I\})\Omega_m\}_{m}$, where only finitely 
many $q^m$ and $q'^m$ are nonvanishing, we have
\begin{eqnarray}
&&\int_{\mathbb{R}}\mathrm{d}t\ 
\langle\psi_{ac}|e^{it(\mathbf{M}-\epsilon)}|\psi'_{ac}\rangle^{ac}\nonumber\\
&=&\sum_{m=1}^\infty\int_{\mathbb{R}}\mathrm{d}t\ 
\langle{\Omega}_m|q^m(\{\tilde{C}_I\})^\dagger\ 
e^{it(\mathbf{M}-\epsilon)}\ 
q'^m(\{\tilde{C}_I\})|\Omega_m\rangle^{ac}\nonumber\\
&=&\sum_{m=1}^\infty\int_{\mathbb{R}}\mathrm{d}t\int_{\mathbb{R}^N}\mathrm{d}\mu^{ac}(\vec{x})\ 
\rho^{ac}_{\Omega_m}(\vec{x})\ e^{it(K_{IJ}x_Ix_J-\epsilon)}\ 
\overline{q^m(\vec{x})}\ q'^m(\vec{x})\nonumber\\
&=&\sum_{m=1}^\infty\int_{\mathbb{R}}\mathrm{d}t\int_{\mathbb{R}^N}\mathrm{d}\mu^{ac}(\lambda,\vec{\xi})\ 
\rho^{ac}_{\Omega_m}(\lambda,\vec{\xi})\ e^{it(\lambda-\epsilon)}\ 
\overline{q^m(\lambda,\vec{\xi})}\ q'^m(\lambda,\vec{\xi})\nonumber\\
&=&\sum_{m=1}^\infty\lim_{g\to0}\int_{\mathbb{R}}\mathrm{d}t\int_{\mathbb{R}^N}\mathrm{d}\mu^{ac}(\lambda,\vec{\xi})\ 
\rho^{ac}_{\Omega_m}(\lambda,\vec{\xi})\ e^{it(\lambda-\epsilon)-|gt|}\ 
\overline{q^m(\lambda,\vec{\xi})}\ q'^m(\lambda,\vec{\xi})\nonumber\\
&=&\sum_{m=1}^\infty\lim_{g\to0}\int_{\mathbb{R}^N}\mathrm{d}\mu^{ac}(\lambda,\vec{\xi})\ 
\frac{2g}{g^2+(\lambda-\epsilon)^2}\ 
\rho^{ac}_{\Omega_m}(\lambda,\vec{\xi})\ 
\overline{q^m(\lambda,\vec{\xi})}\ q'^m(\lambda,\vec{\xi})\nonumber\\
&=&\lim_{g\to0}\int_{\mathbb{R}^N}\mathrm{d}\mu^{ac}(\lambda,\vec{\xi})\ 
\frac{2g}{g^2+(\lambda-\epsilon)^2}\ 
\sum_{m=1}^\infty\rho^{ac}_{\Omega_m}(\epsilon,\vec{\xi})\ 
\overline{q^m(\epsilon,\vec{\xi})}\ q'^m(\epsilon,\vec{\xi})\nonumber
\end{eqnarray}
where in the last step we have used the differentiability of 
$\rho^{ac}_{\Omega_m} q^m q'^m$ in $\mathcal{N}_0$. Since 
$\sum_m\rho^{ac}_{\Omega_m} q^m q'^m$ is continuous on the compact 
region 
$R_{[0,\epsilon]}$ and bounded on the sphere $K_{IJ}x_Ix_J=\epsilon$, 
there 
exist two functions $M_1(\epsilon):=\mathrm{Max}_{\vec{x}\in 
R_{[0,\epsilon]}}\Big[\sum_m\rho^{ac}_{\Omega_m} q^m 
q'^m(\vec{x})-\sum_m\rho^{ac}_{\Omega_m} q^m q'^m(0)\Big]$ and 
$M_2(\epsilon):=\mathrm{Min}_{\vec{x}\in 
R_{[0,\epsilon]}}\Big[\sum_m\rho^{ac}_{\Omega_m} q^m 
q'^m(\vec{x})-\sum_m\rho^{ac}_{\Omega_m} q^m q'^m(0)\Big]$ such that 
$\lim_{\epsilon\to0}M_i(\epsilon)=0$, so 
\begin{eqnarray}
\sum_{m=1}^\infty\rho^{ac}_{\Omega_m}(0) q^m(0) q'^m(0)+M_2(\epsilon) 
\leqslant \sum_{m=1}^\infty\rho^{ac}_{\Omega_m}(\epsilon,\vec{\xi})\ 
\overline{q^m(\epsilon,\vec{\xi})}\ 
q'^m(\epsilon,\vec{\xi})\leqslant\sum_{m=1}^\infty\rho^{ac}_{\Omega_m}(0) 
q^m(0) q'^m(0)+M_1(\epsilon).\nonumber
\end{eqnarray}
Therefore
\begin{eqnarray}
&&\lim_{g\to0}\int_{\mathbb{R}^N}\mathrm{d}\mu^{ac}(\lambda,\vec{\xi})\ 
\frac{2g}{g^2+(\lambda-\epsilon)^2}\Big[\sum_{m=1}^\infty\rho^{ac}_{\Omega_m}(0) 
q^m(0) q'^m(0)+M_2(\epsilon)\Big]\nonumber\\
&\leqslant&\lim_{g\to0}\int_{\mathbb{R}^N}\mathrm{d}\mu^{ac}(\lambda,\vec{\xi})\ 
\frac{2g}{g^2+(\lambda-\epsilon)^2}\sum_{m=1}^\infty\rho^{ac}_{\Omega_m}(\epsilon,\vec{\xi})\ 
\overline{q^m(\epsilon,\vec{\xi})}\ q'^m(\epsilon,\vec{\xi})\nonumber\\
&\leqslant&\lim_{g\to0}\int_{\mathbb{R}^N}\mathrm{d}\mu^{ac}(\lambda,\vec{\xi})\ 
\frac{2g}{g^2+(\lambda-\epsilon)^2}\Big[\sum_{m=1}^\infty\rho^{ac}_{\Omega_m}(0) 
q^m(0) q'^m(0)+M_1(\epsilon)\Big].\nonumber
\end{eqnarray}
So in the limit $\epsilon\to0$
\begin{eqnarray}
&&\lim_{\epsilon\to0}\lim_{g\to0}\int_{\mathbb{R}^N}\mathrm{d}\mu^{ac}(\lambda,\vec{\xi})\ 
\frac{2g}{g^2+(\lambda-\epsilon)^2}\sum_{m=1}^\infty\rho^{ac}_{\Omega_m}(\epsilon,\vec{\xi})\ 
\overline{q^m(\epsilon,\vec{\xi})}\ q'^m(\epsilon,\vec{\xi})\nonumber\\
&=&\lim_{\epsilon\to0}\lim_{g\to0}\int_{\mathbb{R}^N}\mathrm{d}\mu^{ac}(\lambda,\vec{\xi})\ 
\frac{2g}{g^2+(\lambda-\epsilon)^2}\sum_{m=1}^\infty\rho^{ac}_{\Omega_m}(0)\ 
\overline{q^m(0)}\ q'^m(0)\nonumber
\end{eqnarray}
As a result,
\begin{eqnarray}
\langle\tilde{\eta}(\psi)|\tilde{\eta}(\phi)\rangle_\Omega&=&\lim_{\epsilon\rightarrow0}\frac{\int_{\mathbb{R}}\mathrm{d}t\ 
\langle\psi_{ac}|e^{it(\mathbf{M}-\epsilon)}|\phi_{ac}\rangle^{ac}}{\int_{\mathbb{R}}\mathrm{d}t\ 
\langle\Omega_{ac}|e^{it(\mathbf{M}-\epsilon)}|\Omega_{ac}\rangle^{ac}}\nonumber\\
&=&\lim_{\epsilon\rightarrow0}\lim_{g\to0}\frac{\int_{\mathbb{R}^N}\mathrm{d}\mu^{ac}(\lambda,\vec{\xi})\ 
\frac{2g}{g^2+(\lambda-\epsilon)^2}\ 
\sum_{m=1}^\infty\rho^{ac}_{\Omega_m}(0)\ \overline{q^m(0)}\ 
q'^m(0)}{\int_{\mathbb{R}^N}\mathrm{d}\mu^{ac}(\lambda,\vec{\xi})\ 
\frac{2g}{g^2+(\lambda-\epsilon)^2}\ 
\sum_{m=1}^\infty\rho^{ac}_{\Omega_m}(0)\ \overline{f^m(0)}\ 
f^m(0)}\nonumber\\
&=&\frac{\langle\psi_{ac}(0)|\phi_{ac}(0)\rangle^{ac}_{\vec{x}=0}}{\langle\Omega_{ac}(0)|\Omega_{ac}(0)\rangle^{ac}_{\vec{x}=0}}\ 
.\nonumber
\end{eqnarray}
which means that $\mathcal{\tilde{H}}_\Omega$ is isomorphic to 
$\mathcal{H}^{ac}_{\vec{x}=0}$ for a certain choice of $\Omega$. Thus 
the isomorphism between $\mathcal{H}_\Omega$ and 
$\mathcal{\tilde{H}}_\Omega$ has been established.\\
$\Box$

\subsection{An infinite number of Abelianized constraints }

Suppose we have a gauge system with an infinite collection of 
(non-Abelian) irreducible first class constraints $C_I$ 
($I\in\{1,2,\cdots,\infty\}\equiv\aleph_0$). We can still abelianize 
these constraints locally by using the Abelianzation theorem \cite{HT} 
to obtain 
$\tilde{C}_I=R_{IJ}C_J$, such that $\{\tilde{C}_I,\tilde{C}_J\}=0$ which 
is an \emph{infinite} dimensional Abelian constraint algebra. If we 
quantize these abelianized constraints as self-adjoint operators with 
$[\tilde{C}_I,\tilde{C}_J]=0$ on $\mathcal{H}_{Kin}$, a group averaging 
approach can be defined by the Abelian Lie algebra structure of the 
constraint algebra: For each state $\psi$ in a dense subset 
$\mathcal{D}$ of $\mathcal{H}_{Kin}$, a linear functional 
$\eta_\Omega(\psi)$ in the algebraic dual of $\mathcal{D}$ can be 
defined 
formally such that $\forall \phi\in\mathcal{D}$
\begin{eqnarray}
\eta_{\Omega}(\psi)[\phi]:=\lim_{\epsilon_I\rightarrow0}\frac{\int_{\mathbb{R}^\infty}\prod_{I=1}^\infty\mathrm{d}t_I\ 
\langle\psi|\prod_{I=1}^\infty 
e^{it_I(\tilde{C}_I-\epsilon_I)}|\phi\rangle_{Kin}}{\int_{\mathbb{R}^\infty}\prod_{I=1}^\infty\mathrm{d}t_I\ 
\langle\Omega|\prod_{I=1}^\infty 
e^{it_I(\tilde{C}_I-\epsilon_I)}|\Omega\rangle_{Kin}}\nonumber
\end{eqnarray}
where $\Omega\in\mathcal{H}_{Kin}$ is a reference vector. Therefore we 
can define the group averaging inner product on the linear space of 
$\eta_\Omega(\psi)$ via 
$\langle\eta(\psi)|\eta(\phi)\rangle_\Omega:=\eta_{\Omega}(\psi)[\phi]$. 
The resulting Hilbert space is denoted by $\mathcal{H}_\Omega$. 

However, the above definition is formal because 
$\prod_{I=1}^\infty\mathrm{d}t_I$ is not a measure on 
$\mathbb{R}^\infty$. So the above definition for the group averaging 
with a infinite set of Abelianized constraint is not meaningful in 
general. On the other hand, however, the group averaging technique with 
the master constraint $\mathbf{M}$ does not suffer from this problem. 
Because the master 
constraint operator is defined by $\mathbf{M}:=\sum_{I,J\in 
\aleph_0}K_{IJ}\tilde{C}_I\tilde{C}_J$ ($K_{IJ}$ is nondegenerate), 
we can proceed as before: For each state $\psi$ in the dense subset 
$\mathcal{D}$ of $\mathcal{H}_{Kin}$, a linear functional 
$\tilde{\eta}_\Omega(\psi)$ in the algebraic dual of $\mathcal{D}$ can 
be 
defined such that $\forall \phi\in\mathcal{D}$
\begin{eqnarray}
\tilde{\eta}_{\Omega}(\psi)[\phi]:=\lim_{\epsilon\rightarrow0}\frac{\int_{\mathbb{R}}\mathrm{d}t\ 
\langle\psi|e^{it(\mathbf{M}-\epsilon)}|\phi\rangle_{Kin}}{\int_{\mathbb{R}}\mathrm{d}t\ 
\langle\Omega|e^{it(\mathbf{M}-\epsilon)}|\Omega\rangle_{Kin}}\label{FMGA}
\end{eqnarray}
where $\Omega\in\mathcal{H}_{Kin}$ is the reference vector. Therefore we 
can define another group averaging inner product on the linear span of 
the 
$\tilde{\eta}_\Omega(\psi)$ via 
$\langle\tilde{\eta}(\psi)|\tilde{\eta}(\phi)\rangle_\Omega:=\tilde{\eta}_{\Omega}(\psi)[\phi]$. 
The resulting Hilbert space is denoted by $\tilde{\mathcal{H}}_\Omega$. 

So far we see that for the case of an infinite number of constraints, 
the group averaging inner product with Abelianized constraint is a 
priori
ill-defined but the group averaging inner product with a single master 
constraint is well-defined as long as the Master constraint is well 
defined. Thus the question arises how to 
regularise the group averaging inner product for the infinite number of 
constraints such that in the limit as the regulator is removed we obtain 
the group averaging inner product with respect to the Master constraint. 

We solve this problem as follows: Consider arbitrary
finite subsets $W\subset\aleph_0$ and define ($|W|$ is the 
number of elements in $W$) a partial group averaging for the 
$W$-dependent states $\psi_W$, $\phi_W$ and $\O_W$ 
\begin{eqnarray}
\lag\eta_{\Omega_W,W}(\psi_W)\big|\eta_{\Omega_W,W}(\phi_W)\rag_{\O_W}:=\lim_{\eps\to0}\frac{\int_{\mathbb{R}^{|W|}}\prod_{I\in 
W}\mathrm{d}t_I\ \langle\psi_W|\prod_{I\in W} 
e^{it_I(\tilde{C}_I-\epsilon_I)}|\phi_W\rangle_{Kin}}{\int_{\mathbb{R}^{|W|}}\prod_{I\in 
W}\mathrm{d}t_I\ \langle\Omega_W|\prod_{I\in W} 
e^{it_I(\tilde{C}_I-\epsilon_I)}|\Omega_W\rangle_{Kin}},\label{PAGA}
\end{eqnarray}
which is well-defined since $W$ is a finite set. 

Likewise, with the chosen $W\subset\aleph_0$ one can also 
define the partial master constraint operator by truncating the sum 
$\mathbf{M}_W:=\sum_{I,J\in W}K_{IJ}\tilde{C}_I\tilde{C}_J$. Then the 
group averaging can also be defined for this partial master constraint: 
\begin{eqnarray}
\lag\tilde{\eta}_{\Omega_W,W}(\psi_W)\big|\tilde{\eta}_{\Omega_W,W}(\phi_W)\rag_{\O_W}:=\lim_{\epsilon\rightarrow0}\frac{\int_{\mathbb{R}}\mathrm{d}t\ 
\langle\psi_W|e^{it(\mathbf{M}_W-\epsilon)}|\phi_W\rangle_{Kin}}{\int_{\mathbb{R}}\mathrm{d}t\ 
\langle\Omega_W|e^{it(\mathbf{M}_W-\epsilon)}|\Omega_W\rangle_{Kin}}
\label{partial}
\end{eqnarray}
with respect to the same triple of vectors. Now, we have already seen
in the previous section that under the assumptions spelled out in 
Theorem \ref{FA}, the group averaging using the partial master 
constraint is consistent with the partial group averaging using the
Abelianized constraints, that is
\be
\lag\tilde{\eta}_{\Omega_W,W}(\psi_W)\big|\tilde{\eta}_{\Omega_W,W}(\phi_W)\rag_{\O_W}=\lag\eta_{\Omega_W,W}(\psi_W)\big|\eta_{\Omega_W,W}(\phi_W)\rag_{\O_W}
\ee

What we intend to show is 
that the partial group averaging with respect to the 
partial collection of constraints indexed by $W$, Eq.(\ref{PAGA}) 
coincides 
with the 
group averaging using the master constraint Eq.(\ref{FMGA}) when we take 
the limit $W\to\aleph_0$, i.e.
\be
\lim_{W\to\aleph_0}\lag\eta_{\Omega_W,W}(\psi_W)\big|\eta_{\Omega_W,W}(\phi_W)\rag_{\O_W}=\lag\tilde{\eta}(\psi)\big|\tilde{\eta}(\phi)\rag_\Omega
\ee
for suitable sequences of triples $(\psi_W, \phi_W, \O_W)$
such that $(\psi_W, \phi_W, \O_W)\to(\psi, \phi, \O)$ as $W\to\aleph_0$.
Due to \ref{partial}, and if all the assumptions in Theorem \ref{FA} 
hold for all
choices of $W\in\aleph_0$, the task is reduced to prove
\be
\lim_{W\to\aleph_0}\lag\tilde{\eta}_{\Omega_W,W}(\psi_W)\big|\tilde{\eta}_{\Omega_W,W}(\phi_W)\rag_{\O_W}=\lag\tilde{\eta}(\psi)\big|\tilde{\eta}(\phi)\rag_\Omega\label{both}
\ee
for suitable sequences of triples $(\psi_W, \phi_W, 
\O_W)$ converging strongly
to $(\psi, \phi, \O)$ as $W\to\aleph_0$. This is a simplification of the 
problem because now both sides of Eq.(\ref{both}) 
are the group averaging with respect to the master constraints --- one 
side with 
the partial master constraint and the other side with full master 
constraint. Moreover, by Theorem \ref{GADID}, both sides of 
Eq.(\ref{both}) equal to DID physical inner product in their 
absolutely continuous sectors corresponding to their master constraints 
$\mathbf{M}_W$ and $\mathbf{M}$. So in the following we only need to 
show the following relation:
\be
\lim_{W\to\aleph_0}\frac{\langle{\psi}_{W,ac}(0)|{\phi}_{W,ac}(0)\rangle^{ac}_{\lambda_W=0}}{\langle{\Omega}_{W,ac}(0)|{\Omega}_{W,ac}(0)\rangle^{ac}_{\lambda_W=0}}=\frac{\langle{\psi}_{ac}(0)|{\phi}_{ac}(0)\rangle^{ac}_{\lambda=0}}{\langle{\Omega}_{ac}(0)|{\Omega}_{ac}(0)\rangle^{ac}_{\lambda=0}}\label{rel}
\ee
where $\l_W$ and $\l$ denote the spectrum of $\mathbf{M}_W$ and 
$\mathbf{M}$ respectively. Note that in the following we denote the 
absolutely continuous sector of $\mathbf{M}_W$ by $\tilde{\ch}_W^{ac}$, 
and denote the absolutely continuous sector of $\mathbf{M}$ by 
$\tilde{\ch}^{ac}$.

In order to establish the relation Eq.(\ref{rel}), we have to make a 
regularity assumption
on the convergence of the partial master constraint $\mathbf{M}_W$ to
the 
full master constraint $\mathbf{M}$. We need the following theorem 
(See \cite{Simon} for the proof):

\begin{Theorem} \label{strongresolvent} 
Let $\{A_n\}_{n=1}^\infty$ and $A$ be self-adjoint operators and 
$\lim_{n\to\infty}A_n=A$ in the strong resolvent sense (or equivalently, 
$\lim_{n\to\infty}e^{itA_n}=e^{itA}$ strongly for each $t$), then 
$\lim_{n\to\infty}E_n(a,b)=E(a,b)$ strongly provided that 
$a,b\in\mathbb{R}$, $a<b$, and $a,b\notin\sigma^{pp}(A)$. 
\end{Theorem}
We first consider the simple case that all the partial master 
constraints 
$\mathbf{M}_W$ for different $W$ only have absolutely continuous 
spectrum on $\mathcal{H}_{Kin}$, i.e. $\tilde{\ch}_W^{ac}=\ch_{Kin}$ for 
all $W$. Suppose that we have convergence 
$\mathbf{M}=\lim_{W\to\aleph_0}\mathbf{M}_W$ to the full master 
constraint in 
the strong resolvent 
sense, and that the full master constraint also only has absolutely 
continuous spectrum on 
$\mathcal{H}_{Kin}$, i.e. $\tilde{\ch}^{ac}=\ch_{Kin}$. We also assume 
that for $\mathbf{M}$ there exists a minimal set of 
$\Omega_n\in\mathcal{H}_{Kin}$ such that the Radon-Nikodym derivatives 
$\rho_{\Omega_n}$ are continuous at $\lambda=0$ from the right. For any 
$\psi$, $\psi'$ and $\Omega$ in a dense domain $\mathcal{D}$ (defined by 
the condition that  
for any $\psi,\psi'\in \mathcal{D}$, 
$\langle\psi(\lambda)|\psi'(\lambda)\rangle_{\lambda}$ is right 
continuous at $\lambda=0$) of $\mathcal{H}_{Kin}$, we have:
\begin{eqnarray} \label{limit}
\lim_{\lambda\to0^+} 
\frac{\tilde{\mu}_{\psi,\psi'}(\lambda)}{\tilde{\mu}_{\Omega}(\lambda)}
:=\lim_{\lambda\to0^+}
\frac{\langle\psi|E(0,\lambda)|\psi'\rangle_{Kin}}{\langle\Omega|E(0,\lambda)|
\Omega\rangle_{Kin}}
=\frac{\langle\psi(0)|\psi'(0)\rangle_{\lambda=0}}{\langle\Omega(0)
|\Omega(0)\rangle_{\lambda=0}}.\nonumber
\end{eqnarray}
where $E$ is the p.v.m. of $\mathbf{M}$.
 
On the other hand, the projection valued measure for $\mathbf{M}_W$, 
$E_W(a,b)$ equals $E'_W(R_{(a,b)})$ where $R_{(a,b)}$ is the region 
between the spheres $\mathbf{M}_W=a$ and $\mathbf{M}_W=b$ in 
$\mathbb{R}^{|W|}$ and $E'_W=\prod_{I\in W} \;E_I$. Since 
$\mathbf{M}=\lim_{W\to\aleph_0}\mathbf{M}_W$ 
in the strong resolvent sense and $\mathbf{M}$ only has absolutely 
continuous spectrum 
on $\mathcal{H}_{Kin}$, by the above theorem we know that 
$\lim_{W\to\aleph_0}E_W(a,b)=E(a,b)$ strongly.

Given $W\subset\aleph_0$, we can decompose the Hilbert space 
$\mathcal{H}_{Kin}$ with respect to $\{\tilde{C}_I\}_{I\in W}$
\begin{eqnarray}
\mathcal{H}_{Kin}=\oplus_{m=1}^\infty\mathcal{H}_{W,m}\nonumber
\end{eqnarray}
A dense set $\mathcal{D}_W$ consists of the vectors of the form 
$\{q^m(\{\tilde{C}_I\}_{I\in W})\Omega_{W,m}\}_{m=1}^\infty$, where 
$\Omega_{W,m}\in\mathcal{H}_{W,m}$ are $C^\infty$-vectors and $q^m$ are 
polynomials. Suppose we choose any three unit vectors $\psi$, $\psi'$ 
and $\Omega$ in $\mathcal{D}$, as well as any three unit vectors 
$\psi_W$, $\psi'_W$ and $\Omega_W$ in $\mathcal{D}_W$
\begin{eqnarray}
&&\frac{\tilde{\mu}_{\psi,\psi'}(\lambda)}{\tilde{\mu}_{\Omega}(\lambda)}-\frac{\tilde{\mu}_{W,\psi_W,\psi'_W}(\lambda)}{\tilde{\mu}_{W,\Omega_W}(\lambda)}\nonumber\\
&=&\Big[\frac{\tilde{\mu}_{\psi,\psi'}(\lambda)}{\tilde{\mu}_{\Omega}(\lambda)}-\frac{\tilde{\mu}_{W,\psi,\psi'}(\lambda)}{\tilde{\mu}_{W,\Omega}(\lambda)}\Big]+\Big[\frac{\tilde{\mu}_{W,\psi,\psi'}(\lambda)}{\tilde{\mu}_{W,\Omega}(\lambda)}-\frac{\tilde{\mu}_{W,\psi_W,\psi'_W}(\lambda)}{\tilde{\mu}_{W,\Omega_W}(\lambda)}\Big]\nonumber\\
&=&\frac{\tilde{\mu}_{\psi,\psi'}(\lambda)\tilde{\mu}_{W,\Omega}(\lambda)-\tilde{\mu}_{W,\psi,\psi'}(\lambda)\tilde{\mu}_{\Omega}(\lambda)}{\tilde{\mu}_{\Omega}(\lambda)\tilde{\mu}_{W,\Omega}(\lambda)}+\frac{\tilde{\mu}_{W,\psi,\psi'}(\lambda)\tilde{\mu}_{W,\Omega_W}(\lambda)-\tilde{\mu}_{W,\psi_W,\psi'_W}(\lambda)\tilde{\mu}_{W,\Omega}(\lambda)}{\tilde{\mu}_{W,\Omega}(\lambda)\tilde{\mu}_{W,\Omega_W}(\lambda)}\nonumber\\
&=&\frac{\langle\psi|E(\lambda)-E_W(\lambda)|\psi'\rangle_{Kin}\tilde{\mu}_{\Omega}(\lambda)+\tilde{\mu}_{\psi,\psi'}(\lambda)\langle\Omega|E_W(\lambda)-E(\lambda)|\Omega\rangle_{Kin}}{\tilde{\mu}_{\Omega}(\lambda)[\tilde{\mu}_{\Omega}(\lambda)+\langle\Omega|E_W(\lambda)-E(\lambda)|\Omega\rangle_{Kin}]}\nonumber\\
&&+\frac{[\tilde{\mu}_{W,\psi-\psi_W,\psi'}(\lambda)+\tilde{\mu}_{W,\psi_W,\psi'-\psi'_W}(\lambda)]\ 
\tilde{\mu}_{W,\Omega_W}(\lambda)+\tilde{\mu}_{W,\psi_W,\psi'_W}(\lambda)\ 
[\tilde{\mu}_{W,\Omega_W-\Omega,\Omega_W}(\lambda)+\tilde{\mu}_{W,\Omega,\Omega_W-\Omega}(\lambda)]}{[\tilde{\mu}_{\Omega}(\lambda)+\tilde{\mu}_{W,\Omega_W-\Omega,\Omega_W}(\lambda)+\tilde{\mu}_{W,\Omega,\Omega_W-\Omega}(\lambda)+\langle\Omega|E_W(\lambda)-E(\lambda)|\Omega\rangle_{Kin}]\ 
[\tilde{\mu}_{\Omega}(\lambda)+\langle\Omega|E_W(\lambda)-E(\lambda)|\Omega\rangle_{Kin}]}\nonumber
\end{eqnarray}
Since $\mathcal{D}_{W}$ is dense, for given $\epsilon>0$ and three unit 
vectors $\psi$, $\psi'$, $\Omega$ we can find three unit vectors 
$\psi_{W}$, $\psi'_{W}$ and $\Omega_{W}$ in $\mathcal{D}_{W}$ such that 
$||\psi-\psi_{W}||_{Kin}$,$||\psi'-\psi'_{W}||_{Kin}$, 
$||\Omega-\Omega_{W}||_{Kin}<\epsilon$. Using the Schwarz inequality we 
have 
e.g. 
$|\tilde{\mu}_{{W},\psi-\psi_{W},\psi'_{W}}(\lambda)|\leqslant||\psi-\psi_{W}||_{Kin}<\epsilon$ 
and 
$|\tilde{\mu}_{\psi-\psi_{W},\psi'_{W}}(\lambda)|\leqslant||\psi-\psi_{W}||_{Kin}<\epsilon$ 
for all choices of $W$. Next, for given $\lambda>0$, $\epsilon>0$ and 
$\psi$, $\psi'$, $\Omega$, by strong convergence of the p.v.m. $E_W\to 
E$ (pointwise on the spectrum) there exists a 
$W_0(\lambda,\epsilon,\psi,\psi',\Omega)\subset\aleph_0$ such that for 
all $W\supset W_0(\lambda,\epsilon,\psi,\psi',\Omega)$ we have 
$||[E_{W}(\lambda)-E(\lambda)]\psi||_{Kin}$, 
$||[E_{W}(\lambda)-E(\lambda)]\psi'||_{Kin}$, 
$||[E_{W}(\lambda)-E(\lambda)]\Omega||_{Kin}<\epsilon$. It follows e.g. 
$|\langle\psi|E(\lambda)-E_{W}(\lambda)|\psi'\rangle_{Kin}|<\epsilon$. 
Therefore
\begin{eqnarray}
&&\Bigg|\frac{\tilde{\mu}_{\psi,\psi'}(\lambda)}{\tilde{\mu}_{\Omega}(\lambda)}-\frac{\tilde{\mu}_{W,\psi_W,\psi'_W}(\lambda)}{\tilde{\mu}_{W,\Omega_W}(\lambda)}\Bigg|\nonumber\\
&<&\epsilon\times\Bigg\{\frac{\tilde{\mu}_{\Omega}(\lambda)+|\tilde{\mu}_{\psi,\psi'}(\lambda)|}{\tilde{\mu}_{\Omega}(\lambda)[\tilde{\mu}_{\Omega}(\lambda)-\epsilon]}
+2\frac{\tilde{\mu}_{W,\Omega_W}(\lambda)+|\tilde{\mu}_{W,\psi_W,\psi'_W}(\lambda)|}{[\tilde{\mu}_{\Omega}(\lambda)-3\epsilon]\ 
[\tilde{\mu}_{\Omega}(\lambda)-\epsilon]}\Bigg\}\nonumber\\
&\leqslant&\frac{2\epsilon}{\tilde{\mu}_{\Omega}(\lambda)-\epsilon}\Bigg[\frac{1}{\tilde{\mu}_{\Omega}(\lambda)}+\frac{2}{\tilde{\mu}_{\Omega}(\lambda)-3\epsilon}\Bigg]\nonumber\\
&\leqslant&\frac{6\epsilon}{[\tilde{\mu}_{\Omega}(\lambda)-3\epsilon]^2}\label{estimate}
\end{eqnarray}
where we have assumed that given $\lambda>0$ and $\Omega$, we have 
$3\epsilon<\tilde{\mu}_{\Omega}(\lambda)$. So it is clear that given any 
$\delta>0$ we can choose $\epsilon$ such that Eq.(\ref{estimate}) is 
smaller than $\delta$.
 
Furthermore, both 
${\tilde{\mu}_{\psi,\psi'}(\lambda)}/{\tilde{\mu}_{\Omega}(\lambda)}$ 
and 
${\tilde{\mu}_{W,\psi_W,\psi'_W}(\lambda)}/{\tilde{\mu}_{W,\Omega_W}(\lambda)}$ 
are right continuous at $\lambda=0$. Thus we know that for any 
$\epsilon>0$ there exists a $\delta>0$ such that for all 
$0<\lambda<\delta$, 
$|{\tilde{\mu}_{\psi,\psi'}}/{\tilde{\mu}_{\Omega}}(0)-{\tilde{\mu}_{\psi,\psi'}(\lambda)}/{\tilde{\mu}_{\Omega}(\lambda)}|<\epsilon/3$. 
From the last paragraph, we know that for any $\epsilon>0$ there exists 
a $W_0(\lambda,\epsilon,\psi,\psi',\Omega)\subset\aleph_0$ such that for 
all $W\supset W_0(\lambda,\epsilon,\psi,\psi',\Omega)$, we can find the 
unit vectors $\psi_{W}$, $\psi'_{W}$ and $\Omega_{W}$ in 
$\mathcal{D}_{W}$ such that 
$|{\tilde{\mu}_{\psi,\psi'}(\lambda)}/{\tilde{\mu}_{\Omega}(\lambda)}-{\tilde{\mu}_{W,\psi_W,\psi'_W}(\lambda)}/{\tilde{\mu}_{W,\Omega_W}(\lambda)}|<\epsilon/3$. 
If we fix a $W\supset W_0(\lambda,\epsilon,\psi,\psi',\Omega)$ and data 
$\psi_{W}$, $\psi'_{W}$ and $\Omega_{W}$ in $\mathcal{D}_{W}$, there 
exists a $\delta_W$ such that for all $0<\lambda<\min(\delta,\delta_W)$, 
$|{\tilde{\mu}_{W,\psi_W,\psi'_W}(\lambda)}/{\tilde{\mu}_{W,\Omega_W}(\lambda)}-{\tilde{\mu}_{W,\psi_W,\psi'_W}}/{\tilde{\mu}_{W,\Omega_W}}(0)|<\epsilon/3$. 
To summarize: For any 
$\epsilon>0$, there exists a $W$ and three unit vectors $\psi_{W}$, 
$\psi'_{W}$ and $\Omega_{W}$ in $\mathcal{D}_{W}$ such that 
\begin{eqnarray}
&&\Bigg|\frac{\tilde{\mu}_{\psi,\psi'}}{\tilde{\mu}_{\Omega}}(0)-\frac{\tilde{\mu}_{W,\psi_W,\psi'_W}}{\tilde{\mu}_{W,\Omega_W}}(0)\Bigg|\nonumber\\
&\leqslant&\Bigg|\frac{\tilde{\mu}_{\psi,\psi'}}{\tilde{\mu}_{\Omega}}(0)-\frac{\tilde{\mu}_{\psi,\psi'}(\lambda)}{\tilde{\mu}_{\Omega}(\lambda)}\Bigg|+\Bigg|\frac{\tilde{\mu}_{\psi,\psi'}(\lambda)}{\tilde{\mu}_{\Omega}(\lambda)}-\frac{\tilde{\mu}_{W,\psi_W,\psi'_W}(\lambda)}{\tilde{\mu}_{W,\Omega_W}(\lambda)}\Bigg|+\Bigg|\frac{\tilde{\mu}_{W,\psi_W,\psi'_W}(\lambda)}{\tilde{\mu}_{W,\Omega_W}(\lambda)}-\frac{\tilde{\mu}_{W,\psi_W,\psi'_W}}{\tilde{\mu}_{W,\Omega_W}}(0)\Bigg|<\epsilon\nonumber
\end{eqnarray}
for all $0<\lambda<\min(\delta,\delta_W)$, which means that 
${\tilde{\mu}_{W,\psi_W,\psi'_W}}/{\tilde{\mu}_{W,\Omega_W}}(0)$ 
approximates ${\tilde{\mu}_{\psi,\psi'}}/{\tilde{\mu}_{\Omega}}(0)$ as 
closely as we want. Note that for non-unit vectors $\psi$, $\psi'$, 
$\Omega$ we can always re-scale $\psi_{W}$, $\psi'_{W}$ and $\Omega_{W}$ 
such that the approximation still holds. 

Next we consider the case that each $\mathbf{M}_W$ possibly has 
both absolutely continuous and pure point spectrum on $\ch_{Kin}$ and 
that 
$\mathbf{M}=\lim_{W\to\aleph_0}\mathbf{M}_W$ in the strong resolvent 
sense, and that $\mathbf{M}$ not only has absolutely continuous 
spectrum but also pure point spectrum on
$\mathcal{H}_{Kin}$. We denote by 
$\tilde{\mathcal{H}}^{ac}$ the absolutely continuous sector of 
$\mathbf{M}$. On the subspace $\tilde{\mathcal{H}}^{ac}$, the 
restrictions 
$\mathbf{M}|_{\tilde{\mathcal{H}}^{ac}}=\lim_{W\to\aleph_0}\mathbf{M}_W|_{\tilde{\mathcal{H}}^{ac}}$ 
converge also in the strong resolvent sense because by theorem 
\ref{strongresolvent} only the limit $M$ is supposed to have no pure 
point spectrum (which is the case in $\tilde{\mathcal{H}}^{ac}$ by 
definition). Therefore $E_W(\lambda)$ 
converges 
to 
$E(\lambda)$ strongly on $\tilde{\mathcal{H}}^{ac}$. 
Due to $E(\lambda) E(\Delta)=E(\lambda)$ for any $\lambda\in 
[0,\delta]$ we trivially have for any 
$\tilde{\psi}_{ac},\tilde{\psi'}_{ac},\tilde{\Omega}_{ac}
\in\tilde{\mathcal{H}}^{ac}$
\begin{eqnarray}
&&\tilde{\mu}^{ac}_{\tilde{\psi}_{ac},\tilde{\psi'}_{ac}}(\lambda)=\langle\tilde{\psi}_{ac}|E(\lambda)|\tilde{\psi'}_{ac}\rangle_{Kin}=\langle\Psi_{ac}|E(\lambda)|\Psi'_{ac}\rangle^{ac}=\tilde{\mu}^{ac}_{{\Psi}_{ac},{\Psi'}_{ac}}(\lambda)\nonumber\\
&&\tilde{\mu}^{ac}_{\tilde{\Omega}_{ac},\tilde{\Omega}_{ac}}(\lambda)=\langle\tilde{\Omega}_{ac}|E(\lambda)|\tilde{\Omega}_{ac}\rangle_{Kin}=\langle\Omega_{ac}|E(\lambda)|\Omega_{ac}\rangle^{ac}=\tilde{\mu}^{ac}_{{\Omega}_{ac}}(\lambda)\nonumber
\end{eqnarray}
where $\Psi'_{ac}\equiv 
E(\delta)\tilde{\psi'}_{ac}$, $\Omega_{ac}\equiv 
E(\delta)\tilde{\Omega}_{ac}$. 

Let us make the assumption that
there exists a $\delta\in\mathbb{R}^+$, such that for any 
$\tilde{\psi}_{ac}\in\tilde{\mathcal{H}}^{ac}$, $\Psi_{ac}\equiv 
E(\delta)\tilde{\psi}_{ac}$ belongs to all 
$\tilde{\mathcal{H}}_W^{ac}$'s. Then we can repeat the previous 
manipulations carried out for the simple case 
in the Hilbert subspace $\tilde{\mathcal{H}}^{ac}$. Thus, suppose 
$\Psi'_{ac}$, $\Psi_{ac}$ and $\Omega_{ac}$ are unit vectors in 
$\mathcal{H}^{ac}$. Given $\epsilon>0$ we can find three unit vectors 
$\psi_{W,ac}$, $\psi'_{W,ac}$ and $\Omega_{W,ac}$ in a dense domain 
$\mathcal{S}_{W}\subset\tilde{\mathcal{H}}_W^{ac}$, such that 
$||\Psi_{ac}-\psi_{W,ac}||^{ac}$,$||\Psi'_{ac}-\psi'_{W,ac}||^{ac}$, 
$||\Omega_{ac}-\Omega_{W,ac}||^{ac}<\epsilon$. Using again the Schwarz 
inequality 
we have e.g. 
$|\tilde{\mu}_{{W},\Psi_{ac}-\psi_{W,ac},\psi'_{W,ac}}(\lambda)|\leqslant||\Psi_{ac}-\psi_{W,ac}||^{ac}<\epsilon$ 
and 
$|\tilde{\mu}_{\Psi_{ac}-\psi_{W,ac},\psi'_{W,ac}}(\lambda)|\leqslant||\Psi_{ac}-\psi_{W,ac}||^{ac}<\epsilon$ 
for $\lambda\in(0,\delta]$ and all choices of $W$. Next, for given 
$\lambda\in(0,\delta]$, $\epsilon>0$ and $\Psi'_{ac}$, $\Psi_{ac}$, 
$\Omega_{ac}$ there exists a 
$W_0(\lambda,\epsilon,\Psi_{ac},\Psi'_{ac},\Omega_{ac})\subset\aleph_0$ 
such that for all $W\supset 
W_0(\lambda,\epsilon,\Psi_{ac},\Psi'_{ac},\Omega_{ac})$ 
$||[E_{W}(\lambda)-E(\lambda)]\Psi_{ac}||^{ac}$, 
$||[E_{W}(\lambda)-E(\lambda)]\Psi'_{ac}||^{ac}$, 
$||[E_{W}(\lambda)-E(\lambda)]\Omega_{ac}||^{ac}<\epsilon$. It follows 
e.g. 
$|\langle\Psi_{ac}|E(\lambda)-E_{W}(\lambda)|\Psi'_{ac}\rangle^{ac}|<\epsilon$.

Following the previous manipulations we did for the simple case, we can 
see that for any given 
$\tilde{\psi}_{ac}$,$\tilde{\psi'}_{ac}$,$\tilde{\Omega}_{ac}$ in a 
dense domain $\tilde{\mathcal{S}}\subset\tilde{\mathcal{H}}^{ac}$ (for 
any $\tilde{\psi}_{ac}$,$\tilde{\psi'}_{ac}\in\tilde{\mathcal{S}}$, 
$\langle\tilde{\psi}_{ac}(\lambda)|\tilde{\psi'}_{ac}(\lambda)\rangle^{ac}_{\lambda}$ 
is right continuous at $\lambda=0$), and for any $\eps>0$, there exists 
a $W$ and three vectors $\psi_{W,ac}$, $\psi'_{W,ac}$ and 
$\Omega_{W,ac}$ in a dense domain 
$\mathcal{S}_{W}\subset\mathcal{H}_W^{ac}$ such that  
\begin{eqnarray}
&&||E(\delta)\tilde{\psi}_{ac}-\psi_{W,ac}||,||E(\delta)\tilde{\psi'}_{ac}-\psi'_{W,ac}||, 
||E(\delta)\tilde{\O}_{ac}-\Omega_{W,ac}||<\epsilon\nonumber\\
&&\text{and}\ \ \ \ \ \ \ \ \ 
\Bigg|\frac{\tilde{\mu}^{ac}_{\tilde{\psi}_{ac},\tilde{\psi'}_{ac}}}{\tilde{\mu}^{ac}_{\tilde{\Omega}_{ac}}}(0)-\frac{\tilde{\mu}^{ac}_{W,\psi_{W,ac},\psi'_{W,ac}}}{\tilde{\mu}^{ac}_{W,\Omega_{W,ac}}}(0)\Bigg|<\epsilon\nonumber
\end{eqnarray}
Moreover, we can choose a sequence of $\{W_n\}_{n=1}^\infty$ and 
correspondingly three sequences of vectors $\psi_{W_n,ac}$, 
$\psi'_{W_n,ac}$ and $\Omega_{W_n,ac}$ such that for any given 
$\tilde{\psi}_{ac}$,$\tilde{\psi'}_{ac}$,$\tilde{\Omega}_{ac}$ in a 
dense domain $\tilde{\mathcal{S}}\subset\tilde{\mathcal{H}}^{ac}$ 
(defined by the condition that for  
any $\tilde{\psi}_{ac}$,$\tilde{\psi'}_{ac}\in\tilde{\mathcal{S}}$, we 
have that
$\langle\tilde{\psi}_{ac}(\lambda)|\tilde{\psi'}_{ac}(\lambda)
\rangle^{ac}_{\lambda}$ 
is right continuous at $\lambda=0$), and for any $\eps>0$, there exists 
a $N>0$ such that for all $n>N$ 
\begin{eqnarray}
&&||E(\delta)\tilde{\psi}_{ac}-\psi_{W_n,ac}||,||E(\delta)\tilde{\psi'}_{ac}-\psi'_{W_n,ac}||, 
||E(\delta)\tilde{\O}_{ac}-\Omega_{W_n,ac}||<\epsilon\nonumber\\
&&\text{and}\ \ \ \ \ \ \ \ \ 
\Bigg|\frac{\tilde{\mu}^{ac}_{\tilde{\psi}_{ac},\tilde{\psi'}_{ac}}}{\tilde{\mu}^{ac}_{\tilde{\Omega}_{ac}}}(0)-\frac{\tilde{\mu}^{ac}_{W_n,\psi_{W_n,ac},\psi'_{W_n,ac}}}{\tilde{\mu}^{ac}_{W_n,\Omega_{W_n,ac}}}(0)\Bigg|<\epsilon\nonumber
\end{eqnarray}
which means that 
\be
&&\lim_{n\to\infty}\psi_{W_n,ac}=E(\delta)\tilde{\psi}_{ac},\ \ \ \  
\lim_{n\to\infty}\psi'_{W_n,ac}=E(\delta)\tilde{\psi}'_{ac},\ \ \ \ 
\lim_{n\to\infty}\Omega_{W_n,ac}=E(\delta)\tilde{\O}_{ac}\nonumber\\
&&\text{and}\ \ \ \ \ \ \ 
\lim_{n\to\infty}\frac{\tilde{\mu}^{ac}_{W_n,\psi_{W_n,ac},\psi'_{W_n,ac}}}{\tilde{\mu}^{ac}_{W_n,\Omega_{W_n,ac}}}(0)=\frac{\tilde{\mu}^{ac}_{\tilde{\psi}_{ac},\tilde{\psi'}_{ac}}}{\tilde{\mu}^{ac}_{\tilde{\Omega}_{ac}}}(0)=\frac{\langle\tilde{\psi}_{ac}(0)|\tilde{\psi'}_{ac}(0)\rangle^{ac}_{\lambda=0}}{\langle\tilde{\Omega}_{ac}(0)|\tilde{\Omega}_{ac}(0)\rangle^{ac}_{\lambda=0}}
\ee
Note that for a given state $\psi\in \ch_{Kin}$, we should first find 
its 
absolutely continuous component $\psi^{ac}\in\tilde{\ch}^{ac}$ and then 
write 
\be
\psi=E(\delta)\psi^{ac}+\psi_R
\ee
The sequence of vectors converging to $\psi$ is obtained by 
$\psi_{W_n,ac}+\psi_R$ where 
$\lim_{n\to\infty}\psi_{W_n,ac}=E(\delta){\psi}_{ac}$. The remainder
$\psi_R$ is of the form 
$[1-E(\lambda)]\psi_{ac}+\psi_{pp}+\psi_{cs}$ of which the first 
term is projected out by $E(\lambda)$ in the formula \ref{limit}
for the master constraint rigging map already for finite $\lambda\le 
\delta$ and $\psi_{pp},\;\psi_{cs}$ by 
the mechanism of the previous sections as we take the limit $\lambda\to 
0$.  

We summarize the above considerations as the main theorem of 
this 
section:

\begin{Theorem}\label{IA}
Assumptions: 
\begin{itemize}
\item The partial master constraint $\mathbf{M}_W$ converges to 
$\mathbf{M}$ in the strong resolvent sense as $W\to\aleph_0$; 
\item Each $\tilde{C}_I$ satisfies the condition in Theorem \ref{FA} and 
$\mathbf{M}$ satisfies the condition in Theorem \ref{GADID};
\item There exists a $\delta\in\mathbb{R}^+$, such that for any 
${\psi}_{ac}\in\tilde{\mathcal{H}}^{ac}$, $\Psi_{ac}\equiv 
E(\delta){\psi}_{ac}$ belongs to all $\tilde{\mathcal{H}}_W^{ac}$'s
where $\tilde{\mathcal{H}}^{ac},\;\tilde{\mathcal{H}}_W^{ac}$ respectively 
denote the absolutely continuous sector of $\mathbf{M},\;\mathbf{M}_W$
respectively.
\end{itemize}
Then for given states $\psi,\phi,\O$ in a dense domain of $\ch_{Kin}$, 
there exist three sequences $\{\psi_n\}_n,\{\phi_n\}_n,\{\O_n\}_n$ such 
that 
\be
\lim_{n\to\infty}\lt\{\psi_n,\phi_n,\O_n\rt\}=\lt\{\psi,\phi,\O\rt\} \ \ 
\ \ \ \ 
and \ \ \ \ \ \ 
\lim_{n\to\infty}\lag\eta_{\Omega_n,W_n}(\psi_n)\big|\eta_{\Omega_n,W_n}(\phi_n)\rag_{\O_n}=\lag\tilde{\eta}(\psi)\big|\tilde{\eta}(\phi)\rag_\Omega
\ee 
\end{Theorem}

\section{Conclusion and discussion}\label{conc}

In \cite{link} we have tried to sketch how different canonical 
quantisation methods, specifically reduced phase space-, operator
constraint- amd Master constraint quantisation all give rise to the same 
path integral formulation.
In the present paper we have carried out some of the formal steps 
outlined in \cite{link} more carefully, that is, we established a 
tighter relation between DID and group averaging for a single master 
constraint on the one hand and a tighter link between group averaging of 
individidual constraints and Master constraint respectively. Since group 
averaging of individual constraints more or less straightforwardly leads 
to a path integral formulation, we have therefore managed to establish a 
tighter link between DID for a single Master constraint and path 
integrals.

This link rests on assumptions. The mathematical assumptions that we 
have made are rather technical in nature and require rather detailed 
knowledge about the spectral properties of the Master constraints.
They are therefore difficult to verify in concrete situations, however,
they at least caution us that formal manipulations are not granted 
to work out as one would naively expect. Whether they can be weakened 
remains to be seen. On the other hand, since the technical assumptions 
are fulfilled for the examples studied in \cite{DID,DID2} we see that 
they do not restrict us to an empty set of examples. Moreover, even 
if we cannot verify the validity of the assumptions, still it is a 
rather good Ansatz to assume that the DID physical inner product is 
equivalent to a suitable path integral formula. 

The most important physical assumption is that we had to 
assume that the individual constraints form an Abelian algebra of self 
-- adjoint operators. This is consistent with the classical theory 
because any set of first class constraints can be abelianized locally 
in phase space. However, that the constraints be represented without 
anomalies is a strong assumption. Without it, the constraints cannot be 
simultaneously diagonalised on the kinematical Hilbert 
space and then there does not exist a common p.v.m. for all constraints.
In other words, the rigging map then does not produce solutions to all
constraints. To be sure, it is not necessary that the constraint algebra 
be Abelian for group averaging to work. It is sufficient that it is 
a true Lie algebra (structure constants rather than structure functions)
and that there exists a Haar measure on the corresponding gauge group.

However, in the case of GR this is not the case. 
The path integral for GR therefore cannot be derived 
by group averaging of Hamiltonian and spatial diffeomorphism constraint 
operators as envisaged in \cite{Reisenberger} which has already 
been pointed out in the 
second reference of \cite{book}, simply due to the stucture functions.
They cause the Hamiltonian constraint operators not to be self -- 
adjoint which is why a priori they cannot be exponentiated\footnote{The 
exponential of a self -- adjoint operator can be defined via the 
spectral theorem.} and even if they can be defined on analytic vectors
\cite{Simon}, they do not form a Lie algebra\footnote{It is often 
wrongly stated that the Hamiltonian constraint operators \cite{QSD} 
commute. This is wrong. What one means is that the dual action of 
their 
commutators annihilates the 
solutions of the spatial diffeomorphism constraints (which are 
considered as 
distributions on the kinematical Hilbert space). On the kinematical Hilbert 
space they do not 
commute and they do not form a Lie algebra. One can define a Hilbert 
space of solutions to the spatial diffeomorphism constraints. But 
neither is the Hamiltonian constraint defined there (it cannot 
preserve this space) nor is it self -- adjoint. See the second reference 
in \cite{book} for a comprehensive discussion.}. Thus, to derive a path 
integral formula for GR from the canonical theory, we must 
first Abelianize the constraints or one has to use the Master constraint
programme. The general considerations in this paper and the companion 
paper \cite{link} may be considered as a preparation for this.

The consequence of the Abelianisation is that the naive Lebesgue measure 
of a path integral formulation has to be modified by a local measure 
factor. The following sketch may clarify this: Suppose that we have a 
system with only first class constraints $C_I$ and let $\tilde{C}_I$ be 
their local Abelianisation. Then there exists a non singular matrix $M$ 
with $C_I=M_{IJ}\; \tilde{C}_J$. The rigging physical inner product can 
then be formally written as (using the usual skeletonisation techniques)
\begin{eqnarray} \label{sketch}
<\eta[\psi'],\eta[\psi]>_{{\rm phys}} 
&=&
\frac{
\int\; \scrD q\;\scrD p\; \delta[\tilde{C}]\; 
\overline{\psi[q_+]}\;\psi'[q_-]\;exp(i\int\;dt\; p_a\dot{q}^a)
}
{  
\int\; \scrD q\;\scrD p\; \delta[\tilde{C}]\; 
\overline{\Omega[q_+]}\;\Omega[q_-]\;exp(i\int\;dt\; p_a\dot{q}^a)
}
\nonumber\\
&=&
\frac{
\int\; \scrD q\;\scrD p\; \delta[C]\; |\det[M]|\; 
\overline{\psi[q_+]}\;\psi'[q_-]\;exp(i\int\;dt\; p_a\dot{q}^a)
}
{  
\int\; \scrD q\;\scrD p\; \delta[C]\; |\det[M]|\; 
\overline{\Omega[q]_+}\;\Omega[q_-]\;exp(i\int\;dt\; p_a\dot{q}^a)
}
\end{eqnarray}
where the kinematical states are evaluated at boundary configurations 
$q_\mp$ in 
the infinite past and future respectively. 
The appearance of $|\det(M)|$ multiplying the naive Lebesgue measure
$d\mu_L=\scrD q\;\scrD p$ 
is precisely correct and makes sure that the rigging 
inner product above agrees with the one coming from reduced phase space 
quantisation. To see this, notice that the above path integral is 
invariant 
under 
gauge 
transformations canonically generated by the $\tilde{C}_I$ which become 
the identity 
in the infinite past and 
future because this leaves $q_\pm$ invariant, changes the symplectic 
potential $\Theta_L=\int p_a\dot{q}^a dt$ by a total differential which 
vanishes 
at the 
boundaries, as a canonical transformation leaves the Liouville measure 
$d\mu_L$ 
invariant and also the $\tilde{C}$ due to Abelianess. The $\tilde{C}_I$ 
are always of 
the form 
$\pi_I+h_I(\phi^J,Q^A,P_A)$ because one can split the canonical 
pairs $(q^a,p_a)$ into two groups $(\phi^I,\pi_I),\;(Q^A,P_A)$ and solve 
$C_I(q^a,p_a)=0$ in terms of $\pi_I$. The gauge transformation 
$\alpha_\beta=\exp(\beta^I\{\tilde{C}_I,.\})$ acts on the gauge fixing 
condition $G^I=\phi^I-\tau^I$, where $\tau^I=\tau^I(t)$ is an arbitrary 
but fixed configuration, by the shift $\alpha_\beta(G^I)=G^I+\beta^I$.
We therefore trivially have 
\begin{equation}
1=\int\;\scrD \beta \; \delta[\alpha_\beta(G)]
\end{equation}
Thus we can formally run the Fadeev -- Popov argument (we denote by $m$ 
a point 
on 
the phase space)
\begin{eqnarray}
&&
\int\;d\mu_L(m)\;\delta[\tilde{C}(m)]\;\overline{\Omega(q_+(m))}\;
\Omega(q_-(m))\;e^{i\Theta_L(m)}
\nonumber\\
&=&\int\;\scrD \beta\;
\int\;d\mu_L(m)\;\delta[\tilde{C}(m)]\;\delta[(\alpha_\beta \cdot 
G)(m)]\; 
\;\overline{\Omega(q_+(m))}\;
\Omega(q_-(m))\;e^{i\Theta_L(m)}
\nonumber\\
&=&\int\;\scrD \beta\;
\int\;d\mu_L(m)\;\delta[\tilde{C}(m)]\;\delta[G(\alpha_\beta(m))]\; 
\;\overline{\Omega(q_+(m))}\;
\Omega(q_-(m))\;e^{i\Theta_L(m)}
\nonumber\\
&=&\int\;\scrD \beta\;
\int\;d\mu_L(\alpha_\beta(m))\;\delta[\tilde{C}(\alpha_\beta(m))]\;
\delta[G(\alpha_\beta(m))]\; 
\;\overline{\Omega(q_+(\alpha_\beta(m)))}\;
\Omega(q_-(\alpha_\beta(m)))\;e^{i\Theta_L(\alpha_\beta(m))}
\nonumber\\
&=&[\int\;\scrD \beta]\;
\int\;d\mu_L(m)\;\delta[\tilde{C}(m)]\;\delta[G(m)]\; 
\;\overline{\Omega(q_+(m))}\;
\Omega(q_-(m))\;e^{i\Theta_L(m)}
\nonumber\\
\end{eqnarray}
where we have made use of the automorphism property of canonical 
transformations and the invariance properties of the the integrand
at intermediate steps.
Thus the infinite gauge group volume $[\int\;\scrD \beta]$ cancels 
in the fraction (\ref{sketch}) which therefore may be written as 
\begin{eqnarray}
<\eta[\psi'],\eta[\psi]>_{{\rm phys}} 
&=&
\frac{
\int\; \scrD q\;\scrD p\; \delta[C]\;\delta[G]\; |\det[M]|\; 
\overline{\psi[q_+]}\;\psi'[q_-]\;exp(i\int\;dt\; p_a\dot{q}^a)
}
{  
\int\; \scrD q\;\scrD p\; \delta[C]\;\delta[G]\; |\det[M]|\; 
\overline{\Omega[q]_+}\;\Omega[q_-]\;exp(i\int\;dt\; p_a\dot{q}^a)
}
\nonumber\\
&=&
\frac{
\int\; \scrD q\;\scrD p\; \delta[C]\;\delta[G]\; |\det[\{C,G\}]|\; 
\overline{\psi[q_+]}\;\psi'[q_-]\;exp(i\int\;dt\; p_a\dot{q}^a)
}
{  
\int\; \scrD q\;\scrD p\; \delta[C]\;\delta[G]\; |\det[\{C,G\}]|\; 
\overline{\Omega[q]_+}\;\Omega[q_-]\;exp(i\int\;dt\; p_a\dot{q}^a)
}
\end{eqnarray}
which is precisely the well known reduced phase space formula for 
the path integral \cite{HT} which also makes it manifest that the above 
formula is invariant under changes of the gauge fixing condition $G$
at finite times.  

It transpires that, had we not paid attention 
to the fact that we should use a form of the constraints such that they 
form a Lie algebra and such that the rigging map actually maps
to kernel of the constrants, then we would have postulated 
the naive path 
integral in (\ref{sketch}) without the measure factor $\det[M]$ which 
is necessary also in order to be consistent with other well established 
quantisation methods. Since spin foams did start from 
\cite{Reisenberger} where no attention to these subtleties was paid 
and since also current spin foam models based on the Plebanski or Holst
action
\cite{spinfoam2} do not pay attention to these local measure factors,
one may worry whether spin foam models as currently defined actually 
define solutions to the Hamiltonian constraints. In order to investigate 
this question, we have computed in \cite{Engle} the local measure factor 
for Holst gravity because the models in \cite{spinfoam2} are based on 
the Holst action \cite{holst}. The corresponding measure factor, which 
is actually 
more complicated to compute than in the simple situation 
(\ref{sketch}) because Plebanski gravity also contains second 
class constraints, should then be 
incorporated into 
spin foam models which is ongoing work \cite{work}. The measure factor 
destroys the manifest general covariance of the naive measure and one 
may ask whether the corrected measure is invariant at least under 
the gauge transformations generated by the non Abelianised constraints,
that is, the Bergmann -- Komar group \cite{BK}. This is the subject of 
the research conducted in \cite{muxin}.

The Abelianess featured crucially into the proofs of the current paper
in order to establish a link between constraint group averaging and 
master constraint group averaging. However, the Master constraint needs 
not to be defined in terms of Abelianised constraints. Therefore one may
wonder what happens if one tries to define a path integral for Master 
constraint group averaging for the concrete proposal for an LQG master 
constraint in \cite{master} in terms of the original Hamiltonian 
constraints and with a phase space dependent matrix $K$. This analysis 
is carried out in \cite{link2}.

\section*{Acknowledgments}

We thank Jonathan Engle for many insightful discussions.
M.H. gratefully acknowledges the support by International Max Planck 
Research School (IMPRS) and the partial support by NSFC Nos. 10675019 and 10975017, and thanks 
the Perimeter Institute for Theoretical Physics, where 
part of this research took place, for hospitality.

\end{document}